\documentclass{article}
\usepackage{amsmath}
\usepackage{amssymb}
\usepackage{graphicx}
\usepackage{color}

\newcommand{\Slash}[1]{{\ooalign{\hfil/\hfil\crcr$#1$}}}

\newcommand{\wightman}{\tiny \raisebox{-0.5pt}[0pt][0pt]{$\stackrel{ \textstyle >}{\textstyle \rule{0pt}{1.5pt} \smash <}$}}

\newcommand{\be}{\begin{eqnarray}}
\newcommand{\ee}{\end{eqnarray}}
\newcommand{\n}{\nonumber \\}

\begin{document}

\begin{flushright}
{KEK-TH-1724}
\end{flushright}
\vskip 1cm

\begin{center}
{\Large {\bf Coherent Flavour Oscillation \\
and  CP Violating Parameter \\
 in Thermal Resonant Leptogenesis}}%
\vskip 1cm

\renewcommand{\thefootnote}{\fnsymbol{footnote}}
{\large 
Satoshi Iso\footnote[1]{e-mail address: satoshi.iso@kek.jp}
and Kengo Shimada\footnote[2]{e-mail address: skengo@post.kek.jp}
}
\vskip 0.5cm

{\it 
High Energy Accelerator Research Organization (KEK)  \\
and  \\
The Graduate University for Advanced Studies (SOKENDAI), \\
Oho 1-1, Tsukuba, Ibaraki 305-0801, Japan
}
\end{center}

\vskip 1cm
\begin{center}
\begin{bf}
Abstract
\end{bf}
\end{center}
Solving the Kadanoff-Baym (KB) equations in a different method from our previous analysis \cite{ISY}, 
we obtain the CP violating parameter $\varepsilon$ in the thermal resonant leptogenesis
{\it without assuming} smallness of the off-diagonal Yukawa couplings. 
For that purpose, we first derive a kinetic equation for density matrix
of RH neutrinos with almost degenerate masses $M_i \ (i=1,2) \sim M$. 
If the deviation from thermal equilibrium is small, the differential equation is reduced to a 
linear algebraic equation and the density matrix can be solved explicitly in terms of the
time variation of (local) equilibrium distribution function.
The obtained  CP-violating parameter $\varepsilon_i$ is
 proportional to an enhancement factor
$(M^{2}_{i}-M^{2}_{j}) M_i \Gamma_j/ ((M^{2}_{i}-M^{2}_{j})^2 +R_{ij}^2)$ 
with a regulator $R_{ij}=M (\Gamma_i + \Gamma_j)$, 
consistent with the previous analysis \cite{ISY}.
The decay width can be determined systematically  by the 
1PI self-energy of the RH neutrinos in the 2PI formalism.

\newpage
\renewcommand{\thefootnote}{\arabic{footnote}}
\section{Introduction}
Leptogenesis is one of very attractive scenario to explain the baryon number asymmetry \cite{Fukugita:1986hr}
(For review, see \cite{Fong:2013wr}),  
but if the Majorana masses of the right-handed (RH) neutrinos have a hierarchical structure,
the lightest Majorana mass must be heavier than $10^9$ GeV \cite{Davidson:2002qv} in order
to produce sufficient amount of lepton number asymmetry.
The condition can be evaded when Majorana masses are almost degenerate, which is called the 
resonant leptogenesis \cite{Pilaftsis:1997jf}\cite{Pilaftsis:2003gt}\cite{Pilaftsis:2005rv}. 

In  light of the LHC experiment TeV scale leptogenesis has attracted much attention
\cite{Hambye:2001eu}-\cite{Deppisch:2013jxa}. 
Especially, when we try to solve the naturalness problem via the Coleman Weinberg
mechanism in a $B-L$ sector\cite{Iso:2009ss}\cite{Iso:2009nw}, 
$U(1)_{B-L}$ gauge symmetry must be 
spontaneously broken around the TeV scale \cite{Iso:2012jn}
and masses of RH neutrinos are naturally at the same energy scale.
The leptogenesis scale can be much lowered by considering  neutrino 
flavour oscillation  out-of-equilibrium, which is important in the $\nu$MSM scenario 
\cite{Shaposhnikov:2006xi}\cite{Asaka:2006nq}\cite{Asaka:2010kk}\cite{Canetti:2012kh}.
Hence it is becoming more and more important
to treat  coherent flavour oscillation  in a systematic way.

In a conventional approach based on 
the classical Boltzmann equation, the evolution of the phase space distribution functions
of on-shell particles is described and the interactions between particles are taken into account
through the collision terms that comprise the S-matrix elements calculated separately. 
So the conventional classical method is not valid when the quantum coherent oscillation
becomes important such as the flavour oscillations or the resonant leptogenesis. 
Density matrix formalism \cite{Sigl}\cite{Akhmedov:1998qx}
is a multi-flavour generalization 
of the Boltzmann equation and has been applied to neutrino flavour oscillations
\cite{Asaka:2005pn}\cite{Canetti:2010aw}\cite{Gagnon:2010kt}\cite{Asaka:2011wq}.
Another formulation is to use the  Kadanoff-Baym (KB) equation, which is derived from the
 Schwinger-Dyson equation on closed-time-path. 
The approach is very systematic but difficult to solve without introducing various approximations.
It was first applied to the leptogenesis with a hierarchical structure of the Majorana mass 
\cite{Buchmuller:2000nd}, and intensively used in various papers 
 \cite{Anisimov:2010aq}-\cite{Frossard:2013bra}.
 KB equation was applied to the resonant leptogenesis and oscillatory behaviour of 
 lepton asymmetry was discussed
 \cite{De Simone:2007rw}\cite{De Simone:2007pa}\cite{Garbrecht:2011aw}.
The quantum oscillations in the flavored leptogenesis are also discussed
in \cite{Abada:2006fw}\cite{Nardi:2006fx}\cite{Cirigliano:2007hb}\cite{Beneke:2010dz}\cite{Drewes:2012ma}.

In the resonant leptogenesis, CP-asymmetry in the decay of RH neutrinos is generated by
an interference of the tree and the self-energy one-loop diagrams.
The $CP$-violating parameter is given by
\begin{align}
\varepsilon_{i}&\equiv \frac{\Gamma_{N_i \to \ell \phi}-\Gamma_{N_i \to \overline{\ell} \overline{\phi}}}{\Gamma_{N_i \to \ell \phi}+\Gamma_{N_i \to \overline{\ell} \overline{\phi}}}= \sum_{j(\ne i)}\frac{\Im (h^{\dag}h)^{2}_{ij}}{(h^{\dag}h)_{ii}(h^{\dag}h)_{jj}} \frac{ (M^{2}_{i}-M^{2}_{j}) M_{i}\Gamma_{j}}{(M^{2}_{i}-M^{2}_{j})^2+ R_{ij}^2 } 
\label{CPVparameter}
\end{align}
where $h$ is the neutrino Yukawa coupling and $\Gamma_i \simeq (h^{\dag}h)_{ii}M_i /8\pi$ is the decay width of $N_i$.
The resonant enhancement of the CP-violating parameter was discussed in \cite{Flanz:1996fb}, and
systematically  studied in 
\cite{Pilaftsis:2003gt}\cite{Pilaftsis:1997dr}\cite{Pilaftsis:1998pd}.
The regulator was given by $R_{ij}=M_i \Gamma_j$.
If the mass difference is larger than the decay width,
we have $|M^2_i -M^2_j| \gg R_{ij}$, 
and $\varepsilon_i$ is suppressed by $\Gamma_i/M \sim (h^{\dag}h)_{ii}$.
However, in the degenerate case, $|M_i -M_j|\sim \Gamma$ and 
$\varepsilon$ can be enhanced to  ${\cal O}((h^\dagger h)^0)\sim 1$. 
Hence the determination of the regulator $R_{ij}$ is essential for a precise prediction 
of the lepton number asymmetry in the resonant leptogenesis.
The authors \cite{Buchmuller:1997yu} calculated the
resummed propagator of the RH neutrinos and obtained a different regulator
$R_{ij}=|M_i \Gamma_i-M_j \Gamma_j|$. By using their result, the enhancement
factor becomes much larger. The origin of the difference of the regulators 
is discussed in \cite{Anisimov:2005hr} \cite{Rangarajan:1999kt}.

Recently Garny et.al.   \cite{Garny:2011hg} systematically
investigated generation of the lepton asymmetry in the resonant leptogenesis using the 
formulas developed in \cite{Anisimov:2010aq}\cite{Anisimov:2010dk}.
In the investigation, they considered a non-equilibrium initial condition in a time-independent
background and calculated generation of the lepton number asymmetry. 
Starting from the vacuum initial state for the RH neutrinos,
they read  the CP-violating parameter from the generated lepton number asymmetry.
The effective regulator they derived is  $R_{ij}=M_i \Gamma_i+M_j \Gamma_j$, which differs from 
the previous results.

In a previous paper \cite{ISY}, we solved the KB equation in the thermal resonant leptogenesis 
and obtained the same regulator $R_{ij}=M_i \Gamma_i+M_j \Gamma_j$ as above. 
Our derivation is applicable to cases when the background is slowly changing with time
but  valid only when the off-diagonal component of the 
Yukawa couplings are small compared to the diagonal ones
\be
\Re (h^{\dag}h)' <  |M_i-M_j|/M \simeq \Gamma / M \sim (h^{\dag}h)^{d}_{ii} \ .
\label{smallODYukawa}
\ee
For practical purposes, this condition is too strong and it is desirable to extend the analysis
to more general cases with large off-diagonal Yukawa couplings.

The purpose of the paper is to solve the KB equation without assuming 
smallness of the off-diagonal Yukawa couplings (\ref{smallODYukawa}).
In order for it, we first rewrite the KB equation in terms of the density 
matrix of RH neutrinos. Since Majorana fermions have 2 spinor components, 
the density matrix is $2N_F \times 2N_F$ for $N_F$ flavours.
In deriving the kinetic equation for the density matrix, we assume that
deviation of the distribution functions are not very large.
If the condition is satisfied, we reproduce the equation \cite{Sigl}.
Various terms in the equation can be  systematically obtained in the  2PI formalism.
The kinetic equation, which is a differential equation, is reduced to a linear
equation when an inequality $H \ll \Gamma_i$ in (\ref{scales}) between the Hubble parameter 
$H$ and the decay width $\Gamma_i$ of RH neutrino $N_i$ is satisfied. 
Then it is straightforward to obtain the solution of deviation 
of the RH neutrino density matrix from the local equilibrium.
From the off-diagonal component of density matrix, we can read the CP violating parameter $\varepsilon$.
The same CP violating parameter as in \cite{ISY} with the regulator $R_{ij}=M_i \Gamma_i+M_j \Gamma_j$
is obtained.

The paper is organized as follows. In section \ref{sec-KBtoDM}, we derive  kinetic equations of density
matrices starting from the Kadanoff-Baym equations. The derivation is performed under
an assumption that distribution functions are not far from the local equilibrium ones.
But smallness of flavour mixing interactions is {\it not} assumed. Namely, the off-diagonal 
Yukawa couplings are not necessary small compared to the diagonal ones, and 
coherent flavour oscillation is fully taken into account. 
In section \ref{sec-KinDM}, we derive kinetic equations of the 
RH neutrinos and lepton asymmetry in the yield variables. 
In section \ref{sec-KinSol}, we solve the kinetic equations to obtain
the RH neutrino density matrix. From the flavour off-diagonal component, we read the 
CP-violating parameter $\varepsilon$.
We summarize in section \ref{sec-Summary}.
In Appendix A, we explain derivation of  the kinetic term $d_t f$ from KB equation.
Explicit forms of  inverse of matrix ${\cal C}$ are written in Appendix B.
\section{Comparison of time scales}
\setcounter{equation}{0}
We introduce multi-flavour right-handed neutrinos
$\nu _{ R,i}$ where $i$ is the flavour index, $i=1 \cdots N_F.$
In particular we consider a case that 
two RH neutrinos have almost degenerate masses. 
Hence we set  $N_F=2$ in the following.
We write $N_i=\nu_{R,i}+\nu_{R,i}^{c}$.  The Lagrangian is given by
\begin{align}
{\cal L}={\cal L}_{SM}+\frac{1}{2}\overline{N}^{i}(i\Slash{\nabla}-M_i)N^{i}+{\cal L}_{int}\ , \label{L}
\end{align}
\begin{align}
{\cal L}_{int} \equiv -h_{\alpha i}(\overline{\ell}^{\alpha}_{a} \epsilon_{ab} \phi^{*}_{b})P_{R}N^{i}+h^{\dag}_{i\alpha}\overline{N}^{i}P_{L}(\phi_{b} \epsilon_{ba} \ell ^{\alpha}_{a}) \label{Lint}
\end{align}
where  $\alpha ,\beta =1,2,3$ and $a,b=1,2$ are flavor indices of the SM leptons $\ell_a^\alpha$ and
isospin $SU(2)_L$ indices respectively.
$M_i$ is the Majorana mass  of $N_i$ and $h_{i \alpha}$ is the Yukawa coupling of $N^i, \ell_a^\alpha$
and the Higgs $\phi_a$ doublet.
$P_{R(L)}$  are chiral projections on right(left)-handed fermions.
As a concrete model 
we consider the Lagrangian (\ref{Lint}) with only the Yukawa couplings, 
but the following analysis and the results  are not restricted to the specific model: 
we can systematically include other  interactions
such as the  $B-L$ gauge interactions of the RH neutrinos $N_i.$

We  compare  various time (or inverse mass) scales in the model.
First the Hubble parameter  $H$ in the radiation dominant universe is given by
\be
H  \sim 1.66 \sqrt{g_*}\frac{T^{2}}{M_{pl}} \sim \frac{T^2}{10^{18}{\rm GeV}} \ 
\label{HubbleParameter}
\ee
where $T$ is the temperature of the universe.
Thermal masses and decay widths of SM leptons $\ell$ and Higgs $\phi$ are  given by
$m_{\ell,\phi} \sim gT$ and $\Gamma_{\ell, \phi} \sim g^2 T$
where $g$ is the SM gauge coupling. 
When $T$ is lower than 
$g^2 \times 10^{18} GeV$, $\Gamma_{\ell, \phi}$ are larger than $H$. Since we are interested in 
the TeV scale leptogenesis in the present paper, we have the relation
\begin{align}
\Gamma_{\ell ,\phi} \sim g^2 T \gg  H \sim \frac{T^2}{10^{18}{\rm GeV}} \ .
 \label{GammaSMggH}
\end{align}
In type I seasaw model, the decay width of the RH neutrino is given by 
$\Gamma_i \sim (h^{\dag}h)_{ii} M_i/8 \pi$.
The ratio of $\Gamma_i$
to the Hubble parameter (\ref{HubbleParameter})  at temperature $T=M_i$ is 
rewritten 
in terms of the ``effective neutrino mass" $\tilde{m}_i$ as  
(see e.g. \cite{Fong:2013wr})
\be
K_i = \frac{\Gamma_i}{H(M_i)} 
= \frac{\tilde{m}_i}{10^{-3} {\rm eV}}, \ \ \ \tilde{m}_i  \equiv \frac{ (h^\dagger h)_{ii} v^2}{M_i}.
\ee
where $v$ is the scale of the EWSB. 
Hence if we take the Yukawa coupling
so as to $\tilde{m}_i \sim 0.1$ eV, the ratio becomes $K_i \sim 100$.
This corresponds to the strong washout regime.
Hence we have the following inequality among various quantities with mass dimension:
\begin{align}
\Gamma_{\phi}, \Gamma_{\ell} \gg \Gamma_i \gg H \ . \label{scales}
\end{align}
The inequality $\Gamma_{\ell, \phi} \gg \Gamma_i$ is not used in the  analysis 
of the present paper. Hence our results  are still valid
when the RH neutrinos are charged under $B-L$ gauge interaction
and $\Gamma_i$ becomes larger.

\section{From KB  to density matrix evolution \label{sec-KBtoDM}}
\setcounter{equation}{0}
In this section, we derive  an evolution equation of the 
multi-flavour density matrix
of the RH neutrinos $N_i$ \cite{Sigl,Gagnon:2010kt}  
starting from the Kadanoff-Baym equation (see also \cite{Garbrecht:2011aw}). KB equation is 
derived from the Schwinger-Dyson equation on the closed-time-path,
which is a fully
systematic equation of the Green functions in a non-equilibrium setting.
Deriving the kinetic equation for density matrix from the KB equation
makes it clear under what conditions the density matrix equation is obtained
and what kinds of diagrams contribute to various terms in the 
density matrix formalism, especially the resonantly enhanced CP violating parameter
and the {\it decay widths} $\Gamma_i$ contained in the  regulator of $\varepsilon_i.$
\subsection{Green functions}
First we define various Green functions. 
An $ij$-component of Wightman Green functions is defined by
\begin{align}
G_{>}(x,y)_{ij}=  \langle \hat{N_i}(x) \overline{\hat{N_j}}(y) \rangle \ , \ \ 
G_{<}(x,y)_{ij}=  - \langle \overline{\hat{N_j}}(y) \hat{N_i}(x) \rangle \label{defG_{l}} .
\end{align}
The mass $\hat{M}$ and 1PI self-energy function $\Pi$ are also 
$2\times 2$ matrices (besides the spinor structure) 
with the flavour indices $ij$. 
We also define the spectral function  by $G_\rho=i(G_> - G_<)$ and the statistical propagator
by $G_F=(G_> + G_<)/2$.  
The retarded (advanced) Green functions are related to the spectral function  by the relation
\be
G_{R/A}(x,y) &=\pm {\rm \Theta}(\pm (x^0 -y^0))G_{\rho}(x,y).
\ee
For the self-energy function $\Gamma$, we can similarly define various types 
of self-energy functions of
 $R,A,\rho$ and $\wightman$. 
(See Appendix B of \cite{ISY}.)

\begin{figure}[ht]
\begin{center}
\includegraphics[bb=50 500 600 750,clip,width=10cm]{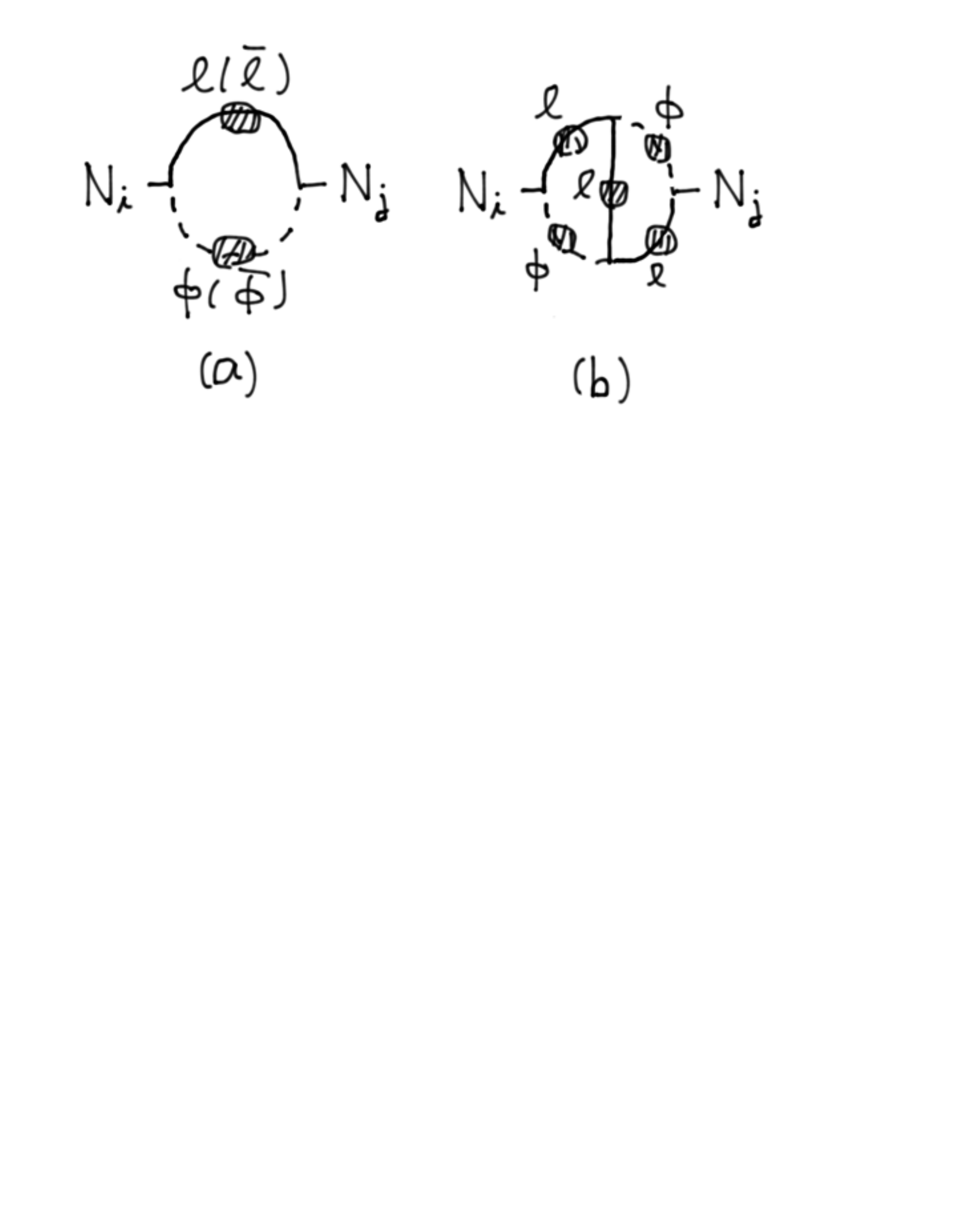}
\caption{
Self-energy diagrams of RH neutrino $N_i$. In the 2PI formalism, each
internal line represents  a full propagator while vertices are given by tree vertices.
Tree-level decay width is generated from the left figure (a).
The right figure (b) gives the so-called 
direct CP violating parameter of the RH neutrino, an interference between 
the tree  and the one-loop vertex corrections.}
\label{Fig2PISE}
\end{center}
\end{figure}
\subsection{Kadanoff-Baym equations}
The Kadanoff-Baym (KB) equation of the RH neutrinos in the expanding universe is given by
\begin{align}
\left( i\gamma^{0}\partial_{x^0}-\frac{{\bold q}\cdot {\boldsymbol \gamma}}{a(x^0)}-\hat{M} \right)G_{\lessgtr}(x^0,y^0)-(\Pi_R * G_{\lessgtr})(x^0,y^0)=(\Pi_{\lessgtr} * G_{A})(x^0,y^0).
\label{KBequationN}
\end{align}
${\bf q}$ is the comoving momentum and $*$ is the convolution in the time coordinate. 
Symbolically we write it as
\be
i G_0^{-1} G_{\lessgtr} - \Pi_R G_{\lessgtr} =\Pi_{\lessgtr} G_A .
\label{KBsymb}
\ee
\begin{figure}[ht]
\begin{center}
\includegraphics[bb=50 500 600 750,clip,width=10cm]{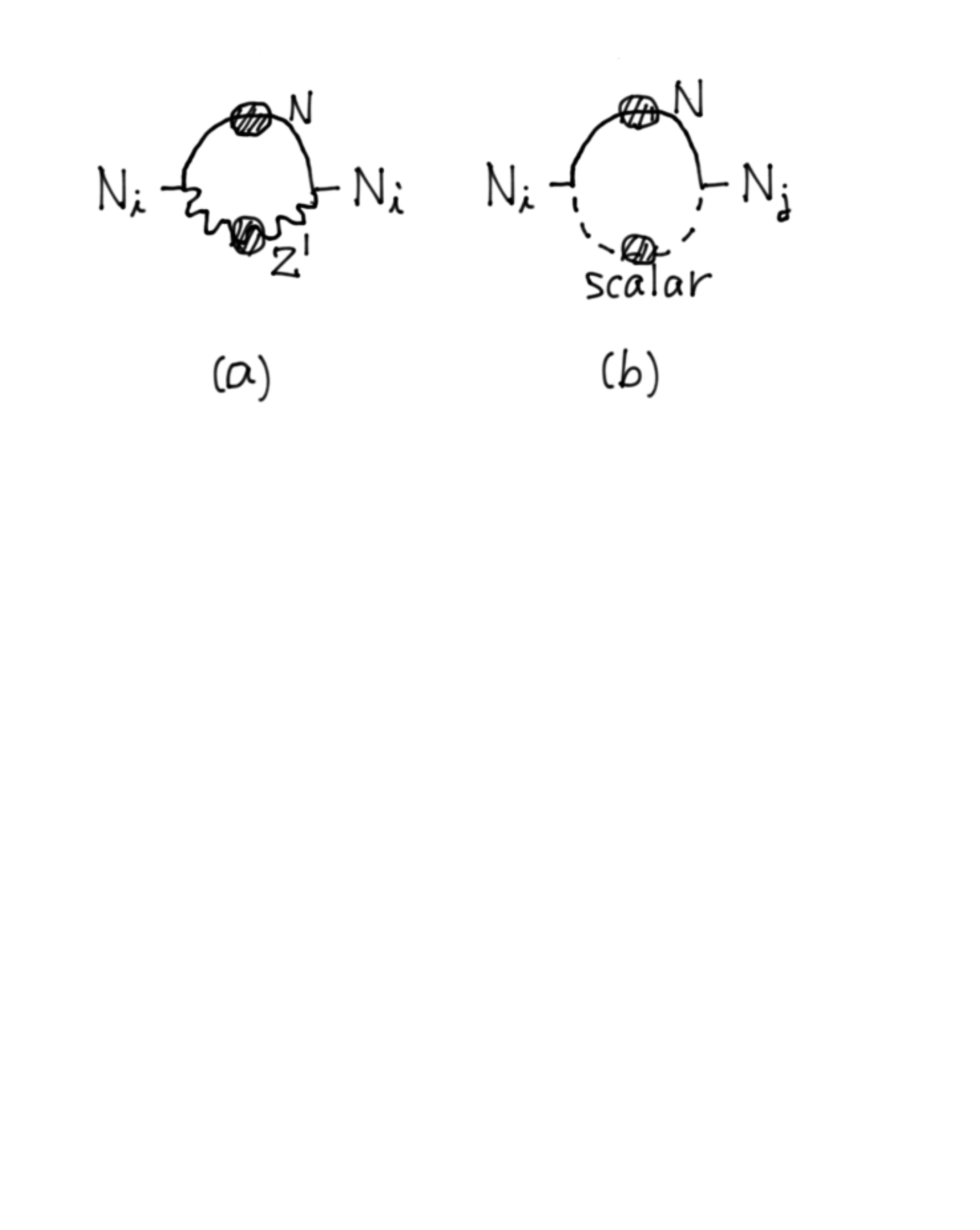}
\caption{
Self-energy diagrams of RH neutrino $N_i$ with $B-L$ gauge 
interaction (a) or with Majorana Yukawa interaction with a
SM singlet scalar field (b). }
\label{Fig2PI-2}
\end{center}
\end{figure}
The 1PI self-energy function $\Pi$ of RH neutrino
is obtained by cutting a (full) propagator of 2PI 
diagrams. In the 2PI formalism, all internal lines  represent full propagators
while vertices are tree.
For more details, see Appendix C, D of \cite{ISY}.
Figure \ref{Fig2PISE} are examples of self-energy diagrams.
In deriving the KB equation, Figure \ref{Fig2PISE}(a) gives the decay width 
at tree level while Figure \ref{Fig2PISE}(b) gives an interference between the 
tree and the one-loop vertex  diagrams \cite{Frossard:2012pc}. Hence the direct CP violating
parameter is contained in fig. \ref{Fig2PISE} (b).
If we include $Z'$ gauge boson or a scalar field coupled with the RH neutrinos,
other self-energy diagrams in Figure \ref{Fig2PI-2} contribute to $\Pi.$

By taking the Fourier transform  with respect to the relative time coordinate
$s=x^0-y^0$,  eq. (\ref{KBequationN}) becomes
\begin{align}
e^{-i\diamondsuit}\left\{ \gamma^{0}q_0-\frac{{\bold q}\cdot {\boldsymbol \gamma}}{a(X)}-\hat{M}-\Pi_R (X;q_0) \right\} \left\{ G_{\lessgtr}(X;q_0)\right\}=e^{-i\diamondsuit}\left\{ \Pi_{\lessgtr}(X;q_0) \right\} \left\{ G_{A}(X;q_0) \right\} .
\label{KBfourierN}
\end{align}
$X=(x^0+y^0)/2$ is the center-of-mass time coordinate.
Here we used the Moyal-Weyl bracket  defined by
\be
e^{-i\diamondsuit} \{f(X;q_0)\}  \{g(X;q_0)\} =
e^{\frac{i}{2}(\partial^f_{q_0} \partial^g_X - \partial^f_X \partial^g_{q_0}  )} f(X;q_0)  g(X;q_0) .
\ee
In the expanding universe with the Hubble parameter $H$,  $X$ derivative is 
often estimated as $\partial_X \sim {\cal O}(H).$
On the other hand,  derivative with respect to the relative momentum $q_0$ is estimated as
$\partial_{q_0} f \sim {\cal O}(1/\Gamma_f)$ where $\Gamma_f$ is the decay width of 
the function $f(X,s) \sim e^{-\Gamma_f s}$. 
In (\ref{KBfourierN}), $\Gamma$ for $G_{*}$
($*= \  \lessgtr, A, R,,,$) is given by the decay width $\Gamma_N$ of the RH neutrinos. 
In the strong washout regime, we have an inequality $H \ll \Gamma_N$. 
Since the dominant contribution to the self-energy $\Pi$ comes from
the diagram in Figure \ref{Fig2PISE} (a),  
$\Gamma$ for $\Pi_{*}$ is given by the decay widths of the charged lepton and Higgs $\Gamma_{l,\phi}$ propagating in the internal lines.
They are much larger than $\Gamma_N$.
An expansion with respect to $\diamondsuit$ is given by
$H/\Gamma_{N,\ell,\phi}$ and hence justified by (\ref{scales}).

Taking up to the first order of the derivative expansion of $\diamondsuit$, 
we have
\begin{align}
&\left( \gamma^{0}q_0-\frac{{\bold q}\cdot {\boldsymbol \gamma}}{a}-\hat{M}-\Pi_R \right) G_{\lessgtr}-i\diamondsuit \left\{ \gamma^{0}q_0-\frac{{\bold q}\cdot {\boldsymbol \gamma}}{a}-\hat{M}-\Pi_R \right\} \left\{ G_{\lessgtr}\right\} \n
& \hspace{5mm}= \Pi_{\lessgtr}  G_{A} -i\diamondsuit \left\{ \Pi_{\lessgtr} \right\} \left\{ G_{A} \right\}
\label{KBderexp}
\end{align}
The   spectral function $G_\rho$ satisfies a similar equation
in which   $\wightman$ ( of $G$ and $\Pi$) is replaced by $\rho$.
\subsection{Green function in the (local) equilibrium \label{GFeq} }
If we drop the derivative term containing $\diamondsuit$, it becomes an equation for the
Green function in the local equilibrium at time $X$;
\begin{align}
&\left( \gamma^{0}q_0-\frac{{\bold q}\cdot {\boldsymbol \gamma}}{a}-\hat{M}-\Pi^{eq}_R \right) G^{eq}_{\lessgtr}= \Pi^{eq}_{\lessgtr}  G^{eq}_{A} . \label{localeq} 
\end{align}
By using the KB equation of the retarded Green fucntion (see eq. (2.11) in \cite{ISY}),
\be
&\left( \gamma^{0}q_0-\frac{{\bold q}\cdot {\boldsymbol \gamma}}{a}-\hat{M}-\Pi^{eq}_R \right) G^{eq}_{R}=-1, 
\ee
eq.(\ref{localeq}) is solved as
\be
G^{eq}_{\lessgtr}= - G^{eq}_{R} \Pi^{eq}_{\lessgtr}  G^{eq}_{A} .
\ee
In the thermal equilibrium at temperature $T$, the Green functions are anti-periodic in the time direction with 
an imaginary period $i \beta=i/T$. Hence
 Fourier transform  satisfies the Kubo-Martin-Schwinger (KMS) relation
\be
G^{(eq)}_{\wightman}(q)=-i\left\{ \begin{matrix} 1-f^{(eq)}(q) \\
-f^{(eq)}(q) \end{matrix} \right\} G^{(eq)}_{\rho}(q), 
\label{KMS}
\ee
where $f^{(eq)}$ is the Fermi distribution function 
$
f^{(eq)}(q_0)=1/(e^{q_0 /T}+1 ).
$
Various properties of the 
equilibrium Green functions are reviewed in section 3.5 in \cite{ISY}.
Especially, as shown in (3.45) in \cite{ISY},
the off-diagonal component of the Wightman functions $G_{\wightman}^{\prime (eq)}(x_0,y_0)$ 
vanishes in the limit of $x_0 \rightarrow y_0.$ It directly follows from the KMS relation together with
the equal-time anti-commutation relation of the fields $N_i$.
When the system is out of equilibrium, it deviates from zero whose imaginary part
 gives the CP violating source for the lepton number asymmetry.

If the system is slightly deviated from the local equilibrium,  KMS relation indicates that the deviation is written as
\be
\delta G_{\wightman}(q) = -i \delta \left\{ \begin{matrix} 1-f(q) \\
-f (q) \end{matrix} \right\} G_{\rho}(q) -i\left\{ \begin{matrix} 1-f(q) \\
-f(q) \end{matrix} \right\} \delta G_{\rho}(q) .
\ee
We then define
\be
\widetilde{\delta G}_{\lessgtr}&\equiv \delta G_{\lessgtr} +i \left[ \begin{matrix} -f \\ 1-f \end{matrix} \right]\delta G_{\rho} = \delta G_{F} +i \left( \frac{1}{2} -f  \right)\delta G_{\rho} .
\ee
which represents a deviation of 
the distribution function  $\widetilde{\delta G}_{\lessgtr} \sim i (\delta f) G_\rho$.
\subsection{KB equation for small deviation from $G_{\lessgtr}^{(eq)}$}
We now derive the KB equation for a small deviation from the local equilibrium.
Taking a variation in (\ref{KBderexp}) and picking
up to the first order terms of $\delta$, we have
\begin{align}
&\left( \gamma^{0}q_0-\frac{{\bold q}\cdot {\boldsymbol \gamma}}{a}-\hat{M}-\Pi^{eq}_R \right) \delta G_{\lessgtr}- \delta \Pi_R G^{eq}_{\lessgtr}
-i\diamondsuit \left\{ \gamma^{0}q_0-\frac{{\bold q}\cdot {\boldsymbol \gamma}}{a}-\hat{M}-\Pi^{eq}_R \right\} \left\{ \delta G_{\lessgtr}\right\} \n
&
-i\diamondsuit \left\{ \gamma^{0}q_0-\frac{{\bold q}\cdot {\boldsymbol \gamma}}{a}-\hat{M}-\Pi^{eq}_R 
-\delta \Pi_R \right\} \left\{ G^{eq}_{\lessgtr}\right\} 
\n
&= \Pi^{eq}_{\lessgtr} \delta G_{A}+\delta \Pi_{\lessgtr}  G^{eq}_{A} 
-i\diamondsuit \left\{ \Pi^{eq}_{\lessgtr} \right\} \left\{ G^{eq}_{A} +\delta G_A \right\} 
-i\diamondsuit \left\{ \delta \Pi_{\lessgtr} \right\} \left\{ G^{eq}_{A} \right\} . \label{deviation}
\end{align}
We can obtain the same equation for $G_{\rho}$ by replacing $\lessgtr$ by $\rho$.
By combining these equations and using the KMS relation, some terms are cancelled and
we have 
\begin{align}
&\left( \gamma^{0}q_0-\frac{{\bold q}\cdot {\boldsymbol \gamma}}{a}-\hat{M}-\Pi^{eq}_R \right) \left( \delta G_{\lessgtr}+i\left[ \begin{matrix} 1-f \\ -f \end{matrix} \right]\delta G_{\rho} \right) \n
&-i\diamondsuit \left\{ \gamma^{0}q_0-\frac{{\bold q}\cdot {\boldsymbol \gamma}}{a}-\hat{M}-\Pi^{eq}_R \right\} \left( \left\{ \delta G_{\lessgtr}\right\} +i\left[ \begin{matrix} 1-f \\ -f \end{matrix} \right]\left\{ \delta G_{\rho} \right\} \right) \n
&-i\diamondsuit \left\{ \gamma^{0}q_0-\frac{{\bold q}\cdot {\boldsymbol \gamma}}{a}-\hat{M}-\Pi^{eq}_R -\delta \Pi_R \right\} \left\{ \left[ \begin{matrix} 1-f \\ -f \end{matrix} \right] \right\} G^{eq}_{\rho} (-i) \n
&=\left( \delta \Pi_{\lessgtr} +i\left[ \begin{matrix} 1-f \\ -f \end{matrix} \right]\delta \Pi_{\rho} \right) G^{eq}_{A} -i(-i)\Pi^{eq}_{\rho}\diamondsuit \left\{ \left[ \begin{matrix} 1-f \\ -f \end{matrix} \right] \right\} \left\{ G_{A}^{eq}+\delta G_{A} \right\} \n
&-i\diamondsuit \left( \left\{ \delta \Pi_{\lessgtr} \right\} +i\left[ \begin{matrix} 1-f \\ -f \end{matrix} \right]\left\{ \delta \Pi_{\rho} \right\} \right) \left\{ G^{eq}_{A} \right\}
\label{W+fRho}
\end{align}
where we defined
\be
\widetilde{\{ \delta G_{\lessgtr}\}}&\equiv \{ \delta G_{\lessgtr} \} + i \left[ \begin{matrix} -f \\ 1-f \end{matrix} \right] \{ \delta G_{\rho} \} =\{ \delta G_{F} \} +i \left( \frac{1}{2} -f \right) \{ \delta G_{\rho} \} . \nonumber
\ee
The deviation from $G_{\lessgtr}^{(eq)}$ occurs due to the expansion of the universe, and hence
$\delta G_{\lessgtr}$ is proportional to the Hubble parameter $H$.
Since the derivative expansion of $\diamondsuit$ is  an expansion of $H$, we can drop terms
containing more than one $\delta$ or $\diamondsuit$ when $H \ll \Gamma_N, \Gamma_{\ell, \phi}$.
Then (\ref{W+fRho}) is simplified as
\begin{align}
&-i\diamondsuit \left\{ \gamma^{0}q_0-\frac{{\bold q}\cdot {\boldsymbol \gamma}}{a}-\hat{M}-\Pi^{eq}_R \right\} \left\{ if  \right\} G^{eq}_{\rho} +i\Pi^{eq}_{\rho}\diamondsuit \left\{ if\right\} \left\{G_{A}^{eq}\right\}\n
&=\widetilde{\delta \Pi_{\lessgtr}}  G^{eq}_{A} 
-\left( \gamma^{0}q_0-\frac{{\bold q}\cdot {\boldsymbol \gamma}}{a}-\hat{M} - \Pi_{R}^{eq}  \right) \widetilde{\delta G_{\lessgtr}} 
\label{left}
\end{align}
Instead of (\ref{KBsymb}), we can start from 
\be
i G_{\lessgtr} G_0^{-1} - G_{\lessgtr} \Pi_A = G_R \Pi_{\lessgtr}
\ee  
and obtain a similar equation to (\ref{left}), 
\begin{align}
&-i G^{eq}_{\rho} \diamondsuit \left\{ if  \right\} \left\{ \gamma^{0}q_0-\frac{{\bold q}\cdot {\boldsymbol \gamma}}{a}-\hat{M}-\Pi^{eq}_A \right\}  +i \diamondsuit \left\{G_{R}^{eq} \right\} \left\{ if\right\} \Pi^{eq}_{\rho} \n
&=G^{eq}_{R} \widetilde{\delta \Pi_{\lessgtr}} 
 - \widetilde{\delta G_{\lessgtr}}\left( \gamma^{0}q_0-\frac{{\bold q}\cdot {\boldsymbol \gamma}}{a}-\hat{M} 
 - \Pi_{A}^{eq} 
 \right) . \label{right}
\end{align} 

By multiplying 
a helicity projection operator with $h=\pm 1$
\be
P_h \equiv \frac{1+h {\bold n}\cdot {\boldsymbol \sigma}}{2}, \ \ {\bf n}=\frac{{\bf q}}{q}, \  \ \sigma^i = \gamma^0 \gamma^i \gamma_5
\ee  
on $\left[ (\ref{left}) - (\ref{right}) \right] $,  and taking trace 
of spinors, we get
\begin{align}
& -i{\rm tr}{\Big [} P_h 
{\Big (}\diamondsuit \left\{ \gamma^{0}q_0-\frac{{\bold q}\cdot {\boldsymbol \gamma}}{a}-\hat{M}-\Pi^{eq}_R \right\} \left\{ if  \right\} G^{eq}_{\rho} 
- \Pi^{eq}_{\rho}
 \diamondsuit \left\{ if\right\} \left\{G_{A}^{eq}\right\} 
\n
&  -
G^{eq}_{\rho} \diamondsuit \left\{ if  \right\} \left\{ \gamma^{0}q_0-\frac{{\bold q}\cdot {\boldsymbol \gamma}}{a}-\hat{M}-\Pi^{eq}_A \right\}  
+ \diamondsuit \left\{G_{R}^{eq} \right\} \left\{ if\right\} \Pi^{eq}_{\rho} {\Big )} {\Big ]} 
\n
=& {\rm tr}{\Big [} P_h 
{\Big (}\left( \hat{M} + \Pi_{H}^{eq} \right) \widetilde{\delta G_{\lessgtr}} - \widetilde{\delta G_{\lessgtr}}\left( \hat{M} + \Pi_{H}^{eq} \right) {\Big )} {\Big ]} \n
+& {\rm tr}{\Big [}  P_h 
{\Big (}\widetilde{\delta \Pi_{\lessgtr}}  G^{eq}_{A} + \frac{1}{2}\Pi_{\rho}^{eq} \widetilde{\delta G_{\lessgtr}} -G^{eq}_{R} \widetilde{\delta \Pi_{\lessgtr}}  +\frac{1}{2} \widetilde{\delta G_{\lessgtr}} \Pi_{\rho}^{eq} {\Big )} {\Big ]} .
\label{frequencyrepr}
\end{align}
where $\Pi_{H}=(\Pi_R+\Pi_A)/2.$

We make the following quasi-particle
ansatz for $\delta \widetilde{G_\lessgtr}$.
In the present paper, we consider a situation that two RH neutrinos have
almost degenerate masses. Hence their poles in the Green function 
can be approximated by a single pole of Breit-Wigner type:
\begin{align}
\widetilde{\delta G_{\lessgtr}}
&\simeq\sum_{h=\pm} i\delta f_{N,h}(q_0,X)  G_{\rho}^{eq} P_h  \n
&\simeq \sum_{h=\pm} (-\delta f_{N,h,q}) \frac{\Gamma_q }{({q_0} -\omega_{q})^2+\Gamma_q^2/4}\frac{\Slash{q}_{+}+M}{2\omega_q} P_h 
\n
&\ +  \sum_{h=\pm}  (-\delta f_{N,h,q}^{*}) \frac{\Gamma_q }{({q_0} +\omega_{q})^2+\Gamma_q^2/4}\frac{\Slash{q}_{-}+M}{2\omega_q} P_h .
\label{QPansatz}
\end{align}
where we set the momentum at on-shell  $q_{\pm \mu}= (\pm \omega_q, -{\bf q})_{\mu}$ and 
\begin{align}
G_{\rho}^{eq}= &\simeq \sum_{{ h}=\pm} \frac{i2q_0 \Gamma_q (\Slash{q}+M)}{({q_0}^2 -\omega_{q}^2)+\omega_{q}^2 \Gamma_q^2} P_{ h}
\label{Lorentz}
\end{align}
is the spectral density of RH neutrino.
Two mass eigenstates are summed in the distribution function $\delta f_N$.
As explained in Section \ref{GFeq}, flavour off-diagonal components of the distribution function
is suppressed by a cancellation of two mass eigenstates. But when the system is out-of-equilibrium,
off-diagonal component of $\delta f_N$ becomes comparable to its diagonal one.

Also note that  hermiticity of Wightman function
 \be
 [G_{<}(q_0,{\bold q})]^{\dag} = \gamma^0 G_{>}(q_0,{\bold q}) \gamma^0
 \ee
together with
 spatial homogeneity and isotropy require the relation $\delta f_{N,h,q}^{\dag} = \delta f_{N,h,q}$.
 Majorana condition 
  \be
  [G_{<}(q_0,{\bold q})]^{C} = C [G_{>}(-q_0,-{\bold q})]^{\rm t} C^{-1} = G_{<}(q_0,{\bold q})
  \ee
   relates the positive and negative frequency parts as in (\ref{QPansatz}).

We then insert the ansatz of $\widetilde{\delta G_\lessgtr}$ of (\ref{QPansatz})
into (\ref{frequencyrepr}) 
and perform $q_0$ integration: $\int_0^\infty dq_0 / 2\pi $ . 
It is dominated near the region 
$ q_0 \sim \omega_{\bold q}=\sqrt{M^2 +|{\bold q}|^2 }$ (see Appendix A), and we get
an evolution equation for the density matrix;
\begin{align}
-id_t  f_{N,h,q} = -[\omega_{qh}^{\rm eff}, \delta f_{N, h,q}] + \frac{{\cal S}}{2} .
\label{DMevoeq}
\end{align}
The density matrix $f_{N,h,q}$ contains an equilibrium part 
 $f^{eq}_{N,h,q}=f_N(\omega_{q}) {\bold 1}_{2\times 2}$
and a deviation from it. 
The derivation of the l.h.s. (the kinetic term $d_t f_{N,h,q}$) is given in Appendix A.
The first term of the r.h.s. in (\ref{frequencyrepr}) gives an effective Hamiltonian,
\begin{align}
\omega_{qh}^{\rm eff} =
{\rm tr}\left\{ \left( \hat{M} +\Pi_{H}^{eq}(q) \right) \frac{\Slash{q}+M}{2\omega_{q}} P_h 
 \right\} ,
\end{align}
while the second term  gives the collision term,
\begin{align}
{\cal S}
&=- {\rm tr}{\Big [} P_h 
{\Big (}\widetilde{\delta \Pi_{\lessgtr}}  G^{eq}_{\rho} -\Pi_{\rho}^{eq} \widetilde{\delta G_{\lessgtr}} +G^{eq}_{\rho} \widetilde{\delta \Pi_{\lessgtr}}  - \widetilde{\delta G_{\lessgtr}} \Pi_{\rho}^{eq} {\Big )} {\Big ]} \n
&=+i {\rm tr}{\Big [} P_h 
{\Big (}\{ \widetilde{\delta \Pi_{>}} ,  G^{eq}_{<} \}+\{  \Pi^{eq}_{>} , \widetilde{\delta G_{<} } \} - \{ \widetilde{\delta \Pi_{<}} , G^{eq}_{>} \} -\{  \Pi^{eq}_{<} , \widetilde{\delta G_{>} } \} {\Big )} {\Big ]} \n
=& +i \left\{ {\rm tr}\left[ P_h 
\frac{\Slash{q}+M}{2\omega_{q}} \delta \Pi_{>}(q) \right] , -f^{eq}_{N,h,q} \right\} 
+i \left\{ {\rm tr}\left[ P_h 
\frac{\Slash{q}+M}{2\omega_{q}} \Pi^{eq}_{>}(q) \right] , -\delta f^{eq}_{N,h,q} \right\} \n
&-i \left\{ {\rm tr}\left[ P_h 
\frac{\Slash{q}+M}{2\omega_{q}} \delta \Pi_{<}(q) \right] , 1-f^{eq}_{N,h,q} \right\} 
-i \left\{ {\rm tr}\left[ P_h 
\frac{\Slash{q}+M}{2\omega_{q}}  \Pi^{eq}_{<}(q) \right] , -\delta f^{eq}_{N,h,q} \right\}
\label{collision}
\end{align}
Here we have used smallness of  the flavour off-diagonal components of 
$G^{eq}_{i\ne j}$ (see discussion after (\ref{KMS})), 
and smallness of flavour dependent thermal corrections to $G_R$.

\section{Kinetic equation for density matrix \label{sec-KinDM}}
\begin{figure}[ht]
\begin{center}
\includegraphics[bb=50 500 600 750,clip,width=10cm]{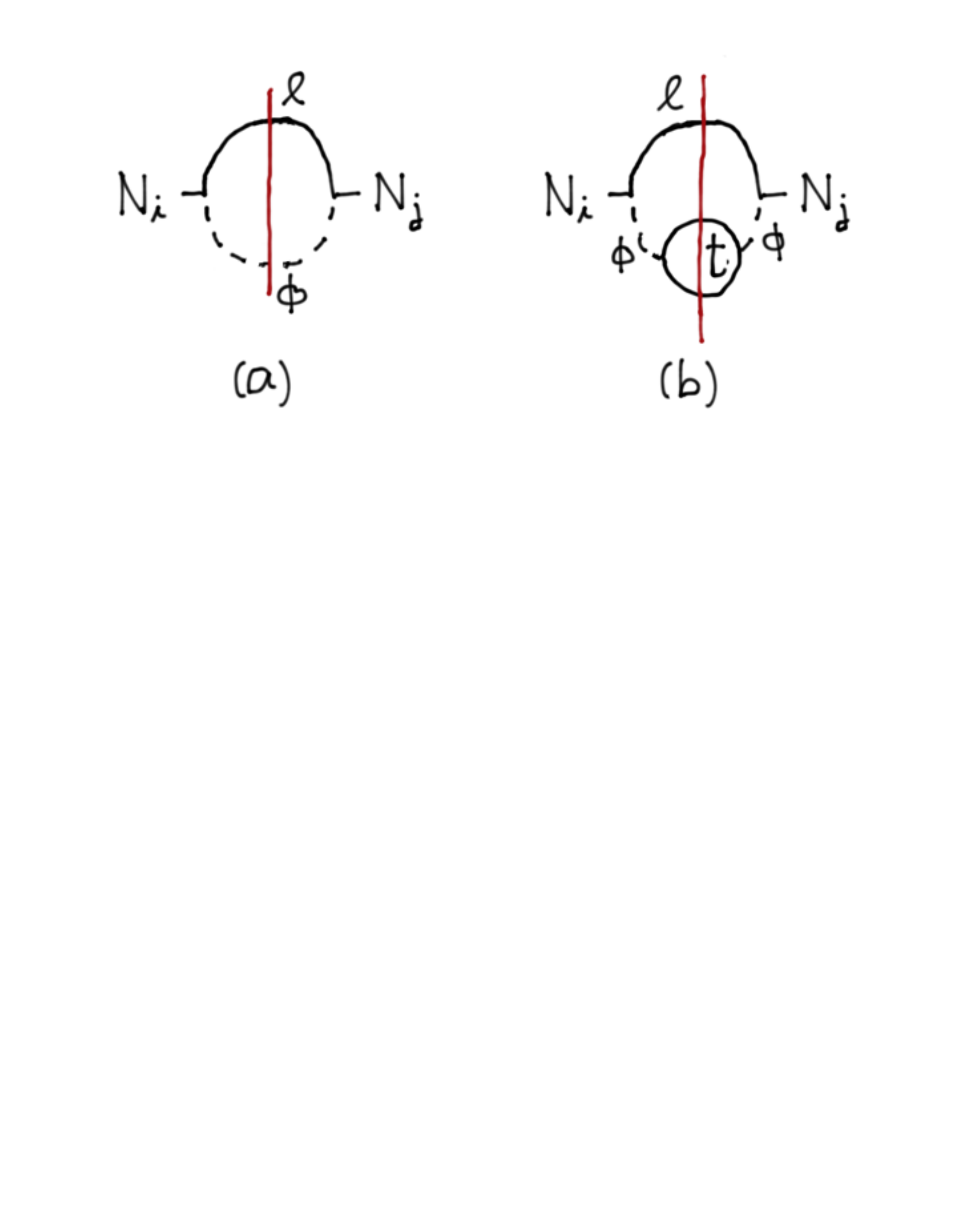}
\caption{
Two dominant contributions to the self-energy diagrams 
of Figure \ref{Fig2PISE} (a). Propagators that cross with 
the cut-line in the middle are put on mass-shell.
 Internal lines are no longer full propagators.
The left figure (a) gives a decay and an inverse-decay term
of RH neutrinos in the KB equation. 
In the right figure (b), 
we consider a loop correction of the Higgs propagator by top quarks. It gives  
scattering terms such as $N+\bar\ell \leftrightarrow t+\bar{Q}$ or
$N +Q \leftrightarrow \ell + t$ in the KB equation\cite{Frossard:2013bra}.
 }
\label{FigCut}
\end{center}
\end{figure}

\setcounter{equation}{0}
In deriving kinetic equations for the density matrix, we need
to make quasi-particle ansatz in (\ref{QPansatz}).
Similar ansatz must be imposed on the internal lines in the
self-energy diagrams $\Pi$ because  distribution functions 
(even when they are matrix-valued) 
are defined only on mass-shell. This is the most subtle point
in the KB approach. In order to take various diagrams
contained in each self-energy diagram in Figure \ref{Fig2PISE},
an often-adopted method is to expand the full propagators
and cut the self-energy diagram into two. Examples are shown in Figure \ref{FigCut}.
On the cut-line, on-shell propagatos are used.
\subsection{Kinetic equation for RH neutrinos}
The collision term (\ref{collision}) is proportional to
\be
 {\rm tr} \left[ P_h \left(
\{\Pi_>, G_< \} -\{\Pi_<, G_> \}
 \right) \right].
\ee
The first term with $G_<$ describes decay (or scattering) of 
RH neutrino (plus other particles ) into others while 
the second term with $G_>$ is an inverse-decay (or inverse scattering).
By expanding the full propagators in the self-energy $\Pi$
 and cutting the diagram into two,
we have various diagrams with on-shell external lines.
External lines are assigned to either incoming or outgoing particles.
If a cut diagram with $G_<$ represents  a scattering process
of $N + i + j \cdots \rightarrow a+b+\cdots$, 
it can be expressed as
\begin{align}
&  - {\rm tr}\left\{ \Pi_{>}(q)( \Slash{q}+M) P_h 
\right\}   
=\sum_{i,..,a,..}\int d\Pi_{a,..,i,..} \gamma_{hij..}^{ab..} f_{i}f_{j}... (1-\eta_{a}f_{a})(1-\eta_{b}f_{b})... 
\label{Ninto}
\end{align}
$\eta_{a,i} =\pm 1$ corresponding to boson or fermion. 
Here the integral measure is defined as
\be
d \Pi_{a,..,i,..} &=& \prod_{a,..,i,..}
\frac{d^3 q_a}{(2 \pi)^3 2 \omega_a} \cdots
\frac{d^3 p_i}{(2 \pi)^3 2 \omega_i} \cdots
\ee
where $q_a$ and $p_i$ are momenta of incoming and outgoing particles.
On the other hand, if a diagram with $G_>$ represents an
inverse scattering process of $a+b+\cdots \rightarrow N+i+j+\cdots$,
it can be expressed as 
\begin{align}
& {\rm tr}\left\{ \Pi_{<}(q)( \Slash{q}+M) P_h 
\right\}  
=\sum_{i,..,a,..}\int d\Pi_{a,..,i,..}  \gamma_{hij..}^{ab..}  (1-\eta_{i}f_{i})(1-\eta_{j}f_{j})...f_{a}f_{b}... 
\label{intoN}
\end{align}
Combining these two contributions, 
the evolution equation for the density matrix $f_{N,h,q}$
(\ref{DMevoeq}) is written  as
\begin{align}
d_{t} f_{N,h,q}=&-i\left[\omega^{\rm eff}_{ q s},f_{N, h, q} \right] \n
&-\frac{1}{2}\frac{1}{2\omega_{ q}}\sum_{i,..,a,..}\int d\Pi_{a,..,i,..}  \{ \gamma_{hij..}^{ab..} ,f_{N,h, q} \} f_{i}f_{j}... (1-\eta_{a}f_{a})(1-\eta_{b}f_{b})... \n
&+\frac{1}{2}\frac{1}{2\omega_{ q}}\sum_{i,..,a,..}\int d\Pi_{a,..,i,..}  \{ \gamma_{hij..}^{ab..} ,(1-f_{N,h, q}) \} (1-\eta_{i}f_{i})(1-\eta_{j}f_{j})...f_{a}f_{b}... 
\label{RHnuDMeq}
\end{align}
In this expression, we combined variations
 as $\Pi=\Pi^{(eq)} + \delta \Pi$ and $f_N=f_N^{(eq)}+\delta f_N$ for 
notational simplicity. 0-th order term of the variation $\delta$ automatically cancels due to the 
detailed balance condition in the equilibrium.

\begin{figure}[ht]
\begin{center}
\includegraphics[bb=50 500 600 750,clip,width=10cm]{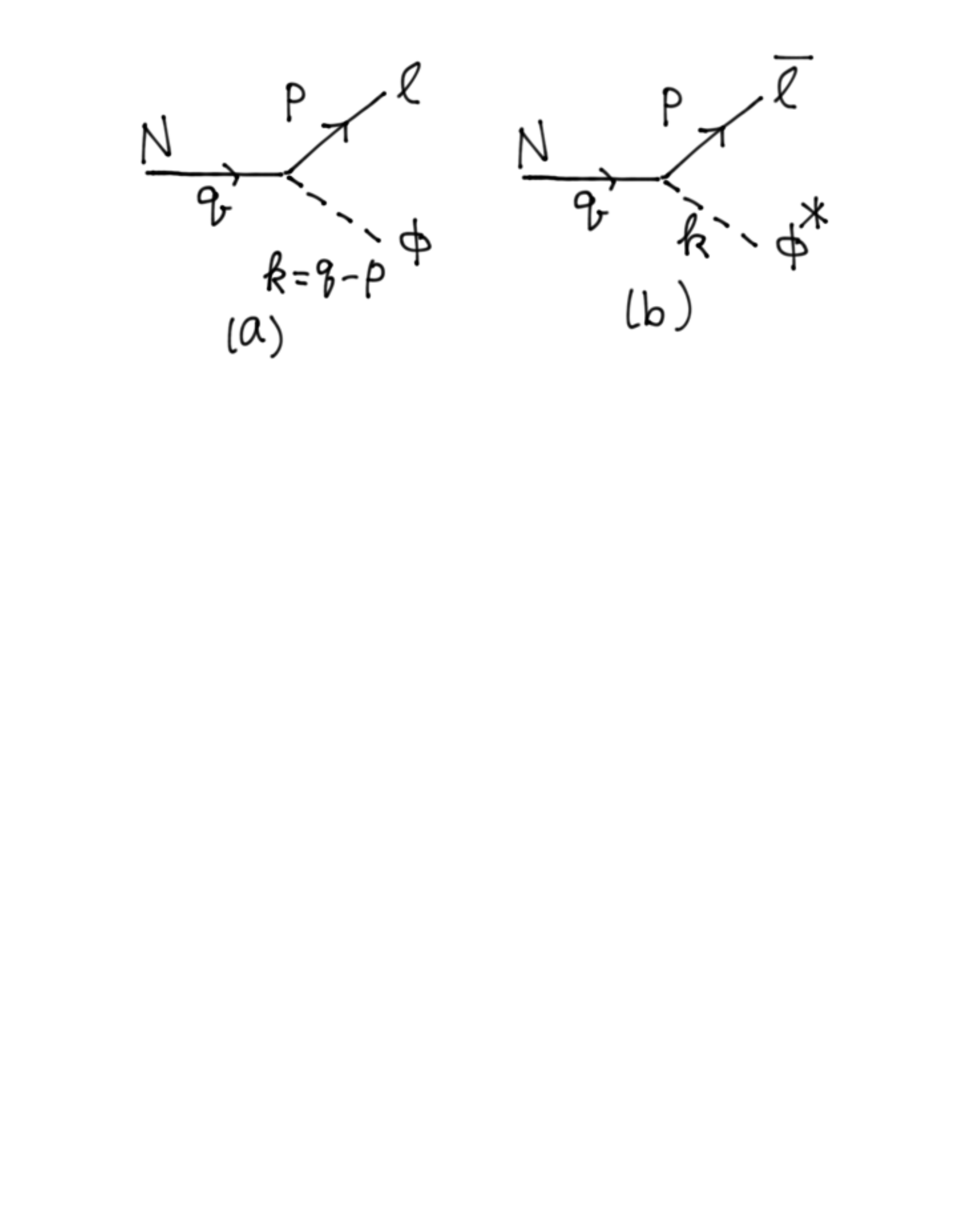}
\caption{
Decay of RH neutrino into $(\ell, \phi)$ and $(\bar\ell, \phi^*)$.
 }
\label{FigVertex}
\end{center}
\end{figure}

Let us now consider a specific diagram of Figure \ref{FigCut} (a).
This diagram  is reduced to the cut diagram of 
Figure \ref{FigVertex} (a).   
Figure \ref{FigVertex} (b) is its conjugate and $N$ decays 
into $(\bar\ell, \phi^*).$
Other diagrams like Figure \ref{FigCut} (b) 
are of higher orders in the Yukawa couplings, and we omit them 
in the following.
From Figure \ref{FigCut}(a) and its conjugate, we have  
\be
\sum_{\alpha}\int d\Pi_{pk} {\big (} \gamma_{h}^{\ell^{\alpha}\phi} (1-f_{\ell^{\alpha} p})(1+f_{\phi k}) + ( \gamma_{-h}^{\ell^{\alpha}\phi})^{*} (1-f_{\overline{\ell}^{\alpha}p})(1+ f_{\overline{\phi}k}) {\big )} 
\ee
for  (\ref{Ninto}), 
and 
\be
\sum_{\alpha}\int d\Pi_{pk} {\big (} \gamma_{h}^{\ell^{\alpha}\phi} f_{\ell^{\alpha}p}f_{\phi k}+ ( \gamma_{-h}^{\ell^{\alpha}\phi})^{*}f_{\overline{\ell}^{\alpha}p}f_{\overline{\phi}k} {\big )} 
\ee
for (\ref{intoN}) 
where  the following relation
\be
\gamma_{h}^{\overline{\ell}^{\alpha}\overline{\phi}} =
(\gamma_{-h}^{\ell^{\alpha}\phi})^{*} 
\ee
is used. 
The decay matrix $\gamma_{h}^{\ell^{\alpha}\phi}$ is given  by
\be
\left( \gamma_{h}^{\ell^{\alpha}\phi} \right)_{ij}
\equiv (h^{\dag}_{i\alpha}h_{\alpha j}) g_w 
\left( q\cdot p -h(\omega_q \frac{{\bold q}\cdot{\bold p}}{|{\bold q}|}
 -\omega_p |{\bold q}| ) \right),
\label{decaymatrix}
\ee
where we have used the relation
\be
{\rm tr} \left( 
( \Slash{q}+M) \frac{1+h {\bold n} \cdot {\boldsymbol \sigma}}{2} \frac{1-\gamma^5}{2} \Slash{p}
 \right) = \left( q\cdot p - h(\omega_q \frac{{\bold q}\cdot{\bold p}}{|{\bold q}|} -\omega_p |{\bold q}| ) \right) .
\ee  
The first term $q\cdot p$ is even under the helicity flip $h \rightarrow -h$,
while the second term is odd. The integral 
\be
\int \frac{d^3 p d^3 k}{2 \omega_p 2 \omega_k} \delta^4(q-p-k) 
(\omega_q \frac{{\bold q}\cdot{\bold p}}{|{\bold q}|} -\omega_p |{\bold q}| ) 
\label{hoddintegral}
\ee
vanishes when  thermal effects of the SM particles, 
namely the thermal mass ($\sim gT$) 
and the statistical factor (Pauli blocking)
of leptons, are neglected.

The kinetic equaction (\ref{RHnuDMeq}) describes an evolution of the density matrix $f_N$ of the RH neutrinos.
Since the equilibrium distribution satisfies the detailed balance condition, the r.h.s. is nonvanishing
only when various quantities are out-of-equilibrium. 
We take a variation of (\ref{RHnuDMeq}) around the equilibrium.
Here note that the relations 
$\delta f_{\ell}=-\delta f_{\overline{\ell}}$、$\delta f_{\phi}=-\delta f_{\overline{\phi}}$
 hold since the SM gauge particles are in thermal
equilibrium and their chemical potentials are vanishing.

In order to solve the kinetic equations,
it is convenient to 
define helicity even and odd combinations 
$\delta f^{even, odd}_{N,q}$ by
\begin{align}
\delta f^{even}_{N, q} \equiv \delta f_{N, +, q} +\delta f_{N, -, q} \ , \ \ \delta f^{odd}_{N, q} \equiv \delta f_{N,+, q} -\delta f_{N,-, q} .
\end{align}
Since helicity operator  ${\bold n}\cdot {\boldsymbol \sigma}$ 
is parity-odd and RH neutrino is invariant under the charge
conjugation, $\delta f^{even, odd}_{N, q}$ are
  CP-even and odd components  respectively;
In terms of these components, 
eq. (\ref{RHnuDMeq}) with the cut-diagram in Figure \ref{FigCut}(a)
can be rewritten as a set of equations 
\begin{align}
& d_{t}(2 f_{N, q}^{eq}+\delta f^{even}_{N, q}) \n
= & -i\left[\frac{\omega^{\rm eff}_{ q +}+\omega^{\rm eff}_{ q -}}{2},\delta f^{even}_{N, q} \right]-i\left[\frac{\omega^{\rm eff}_{ q +}-\omega^{\rm eff}_{ q -}}{2},\delta f^{odd}_{N, q} \right] \n
&-\frac{1}{2}\frac{1}{2\omega_{ q}}\int d\Pi_{pk} \sum_\alpha
\{ \Re ( \gamma_{+}^{\ell^{\alpha}\phi} +\gamma_{-}^{\ell^{\alpha}\phi} ) ,\delta f^{even}_{N, q} \} (1-f_{\ell^{\alpha} p}^{eq}+f_{\phi k}^{eq})  \n
&-\frac{1}{2}\frac{1}{2\omega_{ q}}\int d\Pi_{pk} \sum_\alpha
\{ i \Im ( \gamma_{+}^{\ell^{\alpha}\phi} -\gamma_{-}^{\ell^{\alpha}\phi}) ,\delta f^{odd}_{N q} \} (1-f_{\ell^{\alpha} p}^{eq}+f_{\phi k}^{eq})  \n
&+\frac{1}{2\omega_{ q}}\int d\Pi_{pk} \sum_\alpha
\  i \Im ( \gamma_{+}^{\ell^{\alpha}\phi} +\gamma_{-}^{\ell^{\alpha}\phi} ) {\big (} \delta f_{\ell^{\alpha}p} (f_{\phi k}+f_{N, q}^{eq})+ \delta f_{\phi k} (f_{\ell^{\alpha}p}-f_{N, q}^{eq}) {\big )},
\label{KineticEven}
\end{align}
\begin{align}
& d_{t}(\delta f^{odd}_{N q}) =-i\left[\frac{\omega^{\rm eff}_{ q +}+\omega^{\rm eff}_{ q -}}{2},\delta f^{odd}_{N q} \right]-i\left[\frac{\omega^{\rm eff}_{ q +}-\omega^{\rm eff}_{ q -}}{2},\delta f^{even}_{N q} \right] \n
&-\frac{1}{2}\frac{1}{2\omega_{ q}}\int d\Pi_{pk} \sum_\alpha
\{ \Re (  \gamma_{+}^{\ell^{\alpha}\phi} +\gamma_{-}^{\ell^{\alpha}\phi}) ,\delta f^{odd}_{N q} \} (1-f_{\ell^{\alpha} p}^{eq}+f_{\phi k}^{eq})  \n
&-\frac{1}{2}\frac{1}{2\omega_{ q}}\int d\Pi_{pk} \sum_\alpha
\{ i \Im ( \gamma_{+}^{\ell^{\alpha}\phi} -\gamma_{-}^{\ell^{\alpha}\phi}) ,\delta f^{even}_{N q} \} (1-f_{\ell^{\alpha} p}^{eq}+f_{\phi k}^{eq})  \n
&+\frac{1}{2\omega_{ q}}\int d\Pi_{pk} \sum_\alpha
 \Re ( \gamma_{+}^{\ell^{\alpha}\phi} -\gamma_{-}^{\ell^{\alpha}\phi} ) {\big (} \delta f_{\ell^{\alpha}p} (f_{\phi k}+f_{N, q}^{eq})+ \delta f_{\phi k} (f_{\ell^{\alpha}p}-f_{N, q}^{eq}) {\big )} .
\label{KineticOdd}
\end{align}
If we can neglect the helicity odd part of 
the decay width $\gamma_h^{\ell \phi}$ as discussed in (\ref{hoddintegral})
and the backreaction from lepton asymmetry (the last terms) is 
dropped, these equations for $\delta f^{even}$ and $\delta f^{odd}$
are almost decoupled. Note that the helicity dependent
mass term $(\omega_+  -\omega_-)$ is also negligible if thermal corrections are small.

The dominant source to generate deviations is 
the time variation of the local equilibrium distribution $d_t f^{eq}$,
which is absent in the equation of $\delta f^{odd}$.
Hence in the decoupling limit, it is sufficient to consider only 
the equation for  for $\delta f^{even}$.
In section \ref{SecCPsimple}, we obtain the CP violating parameter under such a condition.

\subsection{Kinetic equation for lepton number }
The evolution equation for the lepton number is similarly obtained 
from the KB equation. Details of the derivation is given in Sec. 2.4 and 2.5 in \cite{ISY}. 
$\alpha$-th flavour lepton number current is defined by
\begin{align}
& \sum_{a} \langle \overline{\hat{\ell}}^{\alpha}_{a}(x) \gamma^{\mu}(x) \ell_{a}^{\alpha} \rangle =-\sum_{a} \left. {\rm tr} \{ \gamma(x) S^{\alpha \alpha}_{aa \lessgtr}(x,y) \} \right|_{y=x} \n
&=-g_w  \left. {\rm tr} \{ \gamma(x) S^{\alpha \alpha}_{\lessgtr}(x,y) \} \right|_{y=x}
\end{align}
where $a$ is an $SU(2)$ isospin index. 
Around $TeV$ scale, the charged Yukawa couplings distinguishing the lepton flavours are in equilibrium and 
the off-diagonal components of lepton flavour density matrix are negligible compared to 
diagonal ones. 
In the second equality, we have assumed that $SU(2)$ isospin 
symmetry is restored.

Since the derivative expansion is an expansion of $H/\Gamma_{\ell, \phi}$, higher order terms
are highly suppressed and we have 
\begin{align}
d_t n_{L^{\alpha}}+3Hn_{L^{\alpha}}& \n
=g_w \int d\Pi_{p}{\Big [}&{\rm tr}\left[ P_{L}\Slash{p} \Sigma^{\alpha \alpha}_{<}(p) \right] (1-f_{\ell^{\alpha} p}) + {\rm tr} \left[ P_{L}\Slash{p} \Sigma^{\alpha \alpha}_{>}(p) \right]f_{\ell^{\alpha} p} \n
-&{\rm tr}\left[ P_{L}\Slash{p} \overline{\Sigma}^{\alpha \alpha}_{<}(p) \right] (1-f_{\overline{\ell}^{\alpha} p}) - {\rm tr} \left[ P_{L}\Slash{p} \overline{\Sigma}^{\alpha \alpha}_{>}(p) \right]f_{\overline{\ell}^{\alpha} p}  {\Big ]} .
\end{align}
$\Sigma$ is the self-energy of the SM lepton $\ell$.
\begin{figure}[ht]
\begin{center}
\includegraphics[bb=50 570 600 750,clip,width=10cm]{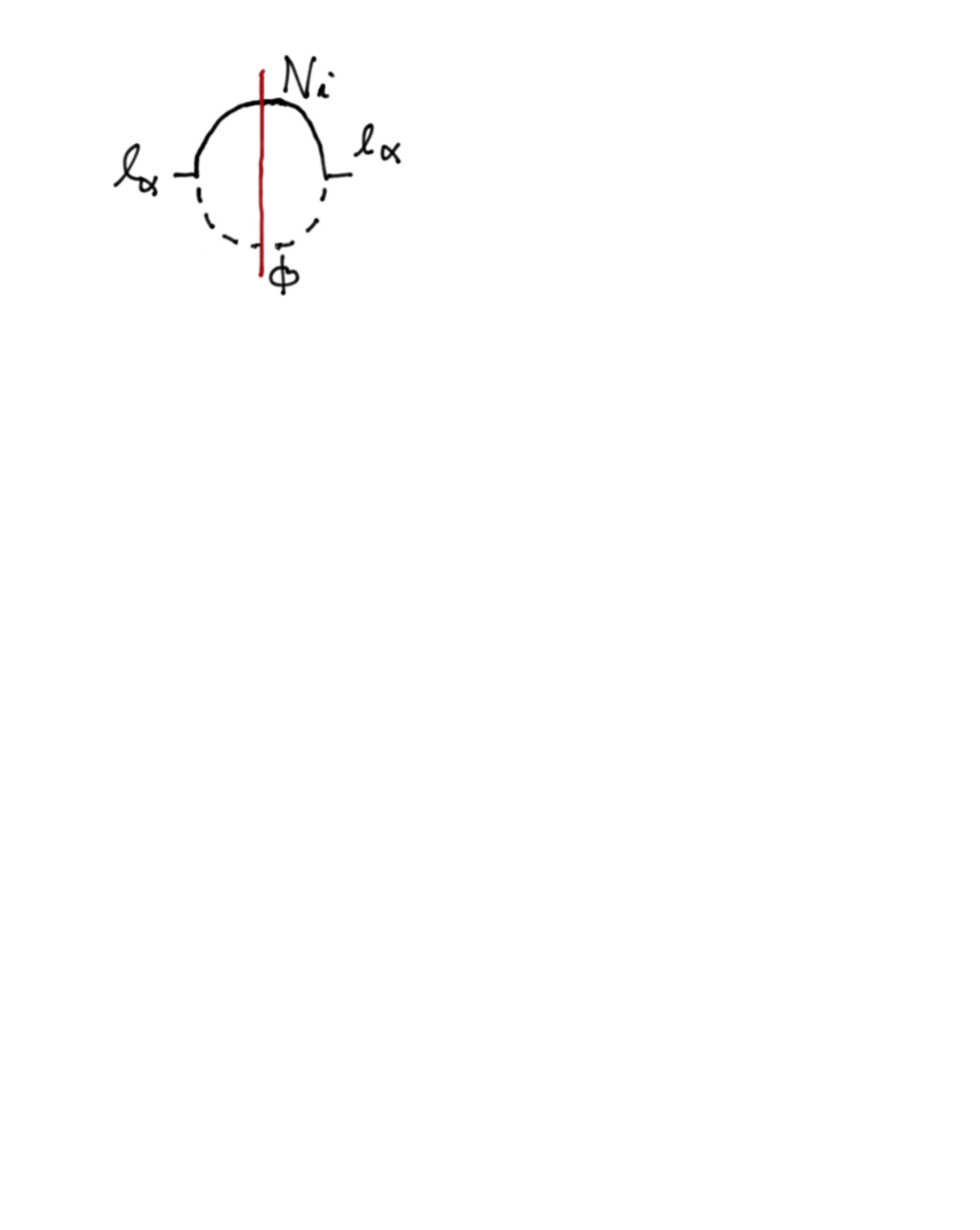}
\caption{
Cutting the self-energy diagram $\Sigma$ of leptons $\ell$.
The cut diagram is the same as Figure \ref{FigVertex} (a).
 }
\label{FigLepSE}
\end{center}
\end{figure}
If we consider, as an example, the Yukawa interaction of $(\ell,\phi,N)$,
the self-energy function for leptons in Figure \ref{FigLepSE}
gives the same 
cut diagram Figure \ref{FigVertex} (a). By using the same 
$\gamma_h^{\ell^\alpha \phi}$ in (\ref{decaymatrix}), 
the kinetic equation is reduced to the following Boltzmann equation;
\begin{align}
d_t n_{L^{\alpha}}+3Hn_{L^{\alpha}}& \n
=\sum_{h} \int d\Pi_{qpk}{\Big [}&{\rm Tr}\left[ \gamma_{h}^{\ell^{\alpha}\phi}\left\{ f_{N, h, q} (1-f_{\ell^{\alpha} p})(1+f_{\phi k}) - (1-f_{N,h, q}) f_{\ell^{\alpha}p}f_{\phi k} \right\} \right] \n
-&{\rm Tr}\left[ (\gamma_{-h}^{\ell^{\alpha}\phi})^{*}\left\{ f_{N,h, q} (1-f_{\overline{\ell}^{\alpha} p})(1+f_{\overline{\phi} k}) - (1-f_{N,h, q}) f_{\overline{\ell}^{\alpha}p}f_{\overline{\phi} k} \right\} \right] {\Big ]} .
\label{KineticLepton}
\end{align}
Here $\rm Tr$ is trace of the RH neutrino flavour.
\subsection{Kinetic equations in terms of Yield variables}
We rewrite the kinetic equations, (\ref{KineticEven}), (\ref{KineticOdd}) and (\ref{KineticLepton}), 
in terms of the Yield variables $Y$ 
defined by 
\begin{align}
Y^{eq}_{N}=\frac{2}{s}  \int \frac{d^3 q}{(2\pi)^3} f^{eq}_{N, q} \ ,\  
Y^{eq}_{\ell^{\alpha}}= \frac{g_w}{s}  \int \frac{d^3 p}{(2\pi)^3} f^{eq}_{\ell^{\alpha} p} \ ,  \
 Y_{L^{\alpha}}= \frac{g_w}{s}  \int \frac{d^3 p}{(2\pi)^3} \left( \delta f_{\ell^{\alpha}} - \delta f_{\overline{\ell}^{\alpha}} \right) .
\end{align}
Here $s$ is the entropy of the universe. 
Note that $Y_N$ is a flavour matrix while $Y_\ell^\alpha$ is a c-number (or $\alpha$-th 
eigenvalue of a diagonal flavour matrix). 
In the following, we consider deviations of distribution functions of RH neutrinos $N_i$ and charged leptons $\ell_\alpha$, 
and other SM particles are assumed to be in the equilibrium distributions. 
We assume $N_i$ and $\ell$ are in the kinematical equilibrium.
Then we can set 
\begin{align}
& \frac{\delta f^{even}_{N, q}}{f^{eq}_{N, q} } = 2\frac{\delta Y_{N}^{even}}{Y^{eq}_{N}} 
\ , \ \frac{\delta f^{odd}_{N, q}}{f^{eq}_{N, q}} = 2 \frac{\delta Y_{N}^{odd}}{Y^{eq}_{N}} \ , \ 
 \frac{\delta f_{\ell^{\alpha}}}{f_{\ell^{\alpha}}^{eq}}=   \frac{Y_{L^{\alpha}}}{2Y^{eq}_{\ell^{\alpha}}}  .
\end{align}    
Since the equations for $\delta Y$ are approximated by coupled linear differential equations,
equations (\ref{KineticEven}), (\ref{KineticOdd}) can be written in a generic form
with matrices 
${\rm H}, \widetilde{\rm H}, \Gamma_N, \widetilde{\Gamma_N}, \Gamma_L, \widetilde{\Gamma_L}$;
\begin{align}
& d_t (Y_{N}^{eq}+\delta Y_{N}^{even})= -i[{\rm H} ,\delta Y_{N}^{even} ] -i [\widetilde{{\rm H}} , \delta Y_{N}^{odd}] \n
& \ \ \  -\frac{1}{2}\{ \Gamma_{N} , \delta Y_{N}^{even} \} -\frac{1}{2}\{ \widetilde{\Gamma}_{N} , \delta Y_{N}^{odd} \} 
+\sum_{\alpha} \Gamma_{L^{\alpha}} Y_{L^{\alpha}} \ ,
 \label{Yieldeven} \\
& d_t (\delta Y_{N}^{odd})=-i[{\rm H} ,\delta Y_{N}^{odd} ] -i [\widetilde{{\rm H}} , \delta Y_{N}^{even}] \n
&\ \ \ -\frac{1}{2}\{ \Gamma_{N} , \delta Y_{N}^{odd} \} -\frac{1}{2}\{ \widetilde{\Gamma}_{N} , \delta Y_{N}^{even} \} 
+\sum_{\alpha} \widetilde{\Gamma}_{L^{\alpha}} Y_{L^{\alpha}} .
\label{Yieldodd}
\end{align}
In the model with only Yukawa interactions, these matrices are given as follows:
\be
{\rm H} &\equiv& \frac{2}{s Y_{N}^{eq}}\int \frac{d^3 q}{(2\pi)^3} f^{eq}_{N, q} \frac{\omega_{+,q}^{\rm eff}+\omega_{-,q}^{\rm eff}}{2} \ , \n
\widetilde{\rm H} &\equiv& \frac{2}{s Y_{N}^{eq}}\int \frac{d^3 q}{(2\pi)^3} f^{eq}_{N, q} \frac{\omega_{+,q}^{\rm eff}-\omega_{-,q}^{\rm eff}}{2} \ ,
\label{H}
\ee
\begin{align}
\Gamma_{N}=\Re \left( \sum_{\alpha}  \Gamma_{\alpha} \right) \ ,\ \ 
\widetilde{\Gamma}_{N}= i \Im \left( \sum_{\alpha}  \widetilde{\Gamma}_{\alpha} \right)
\ ,\ \ \Gamma_{L^{\alpha}}=i \Im [\Gamma_{\alpha}^{W}]
\label{GammaN}
\end{align}
\be
&\widetilde{\Gamma}_{L^{\alpha}} \equiv \frac{1/s}{2Y^{eq}_{\ell^{\alpha}}} \int d\Pi_{qpk}  \Re (  \gamma_{+}^{\ell^{\alpha}\phi} -\gamma_{-}^{\ell^{\alpha}\phi} )  f^{eq}_{\ell^{\alpha}p} (f_{\phi k}+f_{N, q}^{eq}) ,
\label{GammaL}
\ee
where\footnote{The real and imaginary properties of $\Gamma_N$ and $\widetilde{\Gamma_N}$
are valid when we neglect the direct CP violation, an interference between the tree
and one-loop vertex corrections. In the resonant leptogenesis, this approximation is justified.
\label{footnoteRI}
}  
\begin{align}
\Gamma_{\alpha} &\equiv \frac{2}{s Y_{N}^{eq}} \int d\Pi_{qpk} ( \gamma_{+}^{\ell^{\alpha}\phi} +\gamma_{-}^{\ell^{\alpha}\phi})  f^{eq}_{N, q}  (1-f_{\ell^{\alpha} p}^{eq}+f_{\phi k}^{eq})  \\
\widetilde{\Gamma}_{\alpha} &\equiv \frac{2}{s Y_{N}^{eq}} \int d\Pi_{qpk} ( \gamma_{+}^{\ell^{\alpha}\phi} -\gamma_{-}^{\ell^{\alpha}\phi})  f^{eq}_{N,  q}  (1-f_{\ell^{\alpha} p}^{eq}+f_{\phi k}^{eq}) \\
\Gamma_{\alpha}^{W} &\equiv \frac{1/s}{2Y^{eq}_{\ell^{\alpha}}} \int d\Pi_{qpk} (  \gamma_{+}^{\ell^{\alpha}\phi} +\gamma_{-}^{\ell^{\alpha}\phi})  f^{eq}_{\ell^{\alpha}p} (f_{\phi k}+f_{N q}^{eq}) \ .
\end{align}


Similarly the kinetic equation for lepton number (\ref{KineticLepton}) is also rewritten as
\begin{align}
d_t Y_{L^{\alpha}}=&{\rm Tr} \left[ 2 \int d\Pi_{qpk}\ i \Im(  \gamma_{+}^{\ell^{\alpha}\phi} +\gamma_{-}^{\ell^{\alpha}\phi} ) f^{eq}_{N, q} (1-f^{eq}_{\ell^{\alpha} p}+f^{eq}_{\phi k})\frac{\delta Y_{N}^{even}}{s Y_{N}^{eq}} \right] \n
+&{\rm Tr} \left[ 2 \int d\Pi_{qpk}\ \Re(  \gamma_{+}^{\ell^{\alpha}\phi} -\gamma_{-}^{\ell^{\alpha}\phi} )f^{eq}_{N, q} (1-f^{eq}_{\ell^{\alpha} p}+f^{eq}_{\phi k})\frac{\delta Y_{N}^{odd}}{s Y_{N}^{eq}} \right] \n
-& \left[ \int d\Pi_{qpk}\ {\rm Tr}[\Re(  \gamma_{+}^{\ell^{\alpha}\phi} +\gamma_{-}^{\ell^{\alpha}\phi} )]f^{eq}_{N, q} (1+f^{eq}_{\phi k})\frac{Y_{L^{\alpha}}}{s 2Y^{eq}_{\ell^{\alpha}}} \right] \\
=&{\rm Tr} {\Big [} i \Im [\Gamma_{\alpha}]\delta Y_{N}^{even} {\Big ]} 
+{\rm Tr}{\Big [} \Re [\widetilde{\Gamma}_{\alpha}]\delta Y_{N}^{odd} {\Big ]} 
- \Re [\Gamma^{W}_{\alpha}] Y_{L^{\alpha}} \ .
\label{YieldL}
\end{align}
Hence the lepton asymmetry is generated if the r.h.s. is nonvanishing.
CP violating parameter $\varepsilon$ can be read from the equation
by  inserting solutions of  the kinetic equations
for  $\delta Y_N^{even}$  (\ref{Yieldeven}) 
and $\delta Y_N^{odd}$ (\ref{Yieldodd}).
\section{Solution of the kinetic equations \label{sec-KinSol} }
\setcounter{equation}{0}
In order to obtain the CP violating parameter,
we solve the kinetic equations for $\delta Y_N$.
In the derivation of the kinetic equation from the KB equation, 
we assumed that the system is not far from the local equilibrium at each
time of the expanding universe. But  smallness of the off-diagonal Yukawa coupling is {\it not} assumed,
and the coherent flavour oscillation is fully taken into account.
Since the deviation from local equilibrium is caused by the Hubble expansion, both of 
$\delta$ and $\partial_t$ are proportional to the Hubble parameter $H$.
Hence we can set 
\begin{align}
d_t (\delta Y_{N}^{even}) \simeq 0 \ ,\ \ d_t (\delta Y_{N}^{odd}) \simeq 0
\end{align}
in the l.h.s. of Eq. (\ref{Yieldeven}), (\ref{Yieldodd}) under the condition $H \ll \Gamma_i \ll \Gamma_{\ell, \phi}.$

\subsection{Formal solution of $\delta Y_N$}
In the two-flavour case, $Y_N$, $H$, $\Gamma_N$  etc. are  $2 \times 2$ matrices. 
We here express
a $2\times 2$ matrix $A$  as $A=\sum_{a=0}^3  [A]^a \sigma^a$
where  $\sigma^0=1_{2\times 2}$  and  $\sigma^i$ ($i=1,2,3$) is the Pauli matrix.
Then Eqs. (\ref{Yieldeven}), (\ref{Yieldodd}) are rewritten as
\begin{align}
[d_t Y_{N}^{eq}]^{a}&=C^{ab} [ \delta Y_{N}^{even}]^{b} + \widetilde{C}^{ab} [\delta Y_{N}^{odd}]^{b}+[\mu]^{a} \n
0&=C^{ab} [\delta Y_{N}^{odd} ]^{b} + \widetilde{C}^{ab} [ \delta Y_{N}^{even} ]^{b}+[ \widetilde{\mu} ]^{a}
\label{linear}
\end{align}
where
\begin{align}
C^{ab}\equiv&
-\left( \delta^{ab}[\Gamma_{N}]^{0}+\delta^{a}_{0}\delta^{b}_{i}[\Gamma_{N}]^{i}+\delta^{a}_{i}\delta^{b}_{0}[\Gamma_{N}]^{i} +2\delta^{a}_{i}\delta^{b}_{j}\epsilon^{ijk}[{\rm H}]^{k} \right),  \n
\widetilde{C}^{ab}\equiv& 
- \left( \delta^{ab}[\widetilde{\Gamma}_{N}]^{0}+\delta^{a}_{0}\delta^{b}_{i}
[\widetilde{\Gamma}_{N}]^{i}+\delta^{a}_{i}\delta^{b}_{0}[\widetilde{\Gamma}_{N}]^{i} +2\delta^{a}_{i}\delta^{b}_{j}\epsilon^{ijk}[\widetilde{{\rm H}}]^{k} \right), \n
\ [\mu]^{a} \equiv &  \sum_{\alpha} [\Gamma_{L^{\alpha}}]^{a} Y_{L^{\alpha}} \ ,  \ \ 
\ [\widetilde{\mu}]^{a} \equiv  \sum_{\alpha} [\widetilde{\Gamma}_{L^{\alpha}}]^{a} Y_{L^{\alpha}} \ .
 \end{align}
The  Yield density matrix $Y_N^{(eq)}$ in  equilibrium has an $a=0$ component  
only\footnote{ This statemet is correct only when we use the equilibrium distribution function for $f_N.$ If we take higher order terms (the second term
 of eq. (\ref{sharpspectrum})) into account, the off-diagonal components appear
and the following solutions of $\delta Y$ become more complicated. }
\begin{align}
[d_t Y_{N}^{eq}]^{a} = \delta^{a}_{0} (d_t Y_{N}^{eq}) \ .
\end{align}
From (\ref{H}), (\ref{GammaN}) and (\ref{GammaL}), 
${\rm H}, \Gamma_N$  and $\widetilde{\Gamma}_L$ (hence $\widetilde{\mu}$) are real matrices.
Hence $[\Gamma_{N}]^a,[{\rm H}]^{a},[\widetilde{\mu}]^{a}$
do not have an $a=2$ component.
On the other hand, 
$[\widetilde{\Gamma}_{N}]^a,[\widetilde{{\rm H}}]^{a},[\mu]^{a}$ have only an
$a=2$ component since they are imaginary matrices\footnote{Flavour covariance is 
explicitly broken by setting the Majorana mass matrix of the RH neutrinos diagonal with eigenvalues
$M_1, M_2$.}. 

The equations (\ref{linear})
are linear equations with respect to $\delta Y_N$ and can be solved in terms of 
the time-variation of the local equilibrium distribution $d_t Y_N^{(eq)}$ and the lepton asymmetry 
$\mu, \widetilde{\mu}$ as
\begin{align}
\left( \begin{matrix} [\delta Y_{N}^{even}] \\ [\delta Y_{N}^{odd}] \end{matrix} \right) &= {\bold C}^{-1} \left( \begin{matrix} [d_t Y_{N}^{eq}] -[\mu] \\ - [\widetilde{\mu}] \end{matrix} \right) , \ \ \
{\bold C}\equiv \left( \begin{matrix} C & \widetilde{C} \\ \widetilde{C} & C \end{matrix} \right)
\end{align}
In the expanding universe, the deviation of RH neutrino number densities from equilibrium $\delta Y_N$ is first generated and then
lepton asymmetry $Y_L$ is generated by the flavour oscillation and decay.
Here we neglect backreaction from $Y_{L}$ and evaluate the deviation of RH neutrino density directly caused by the expansion of universe. Setting $\widetilde\mu=0$,  $\delta Y_{N}$ is solved as
\begin{align}
[\delta Y_{N}^{even}]^{a}  &= ({\cal C}^{-1})^{ab} [d_t Y_{N}^{eq}]^{b} = ({\cal C}^{-1})^{a0} \times d_t Y_{N}^{eq} \ , \notag \\
[\delta Y_{N}^{odd}]^{a}  &= (\widetilde{\cal C}^{-1})^{ab} [d_t Y_{N}^{eq}]^{b} = (\widetilde{\cal C}^{-1})^{a0} \times d_t Y_{N}^{eq} \label{deltaYieldN} 
\end{align}
where
\begin{align}
{\bold C}^{-1}\equiv \left( \begin{matrix} {\cal C}^{-1} & \widetilde{\cal C}^{-1} \\ \widetilde{\cal C}^{-1} & {\cal C}^{-1} \end{matrix} \right) \ .
\end{align}
Components in the 0-th column of ${\bold C}^{-1}$ are given by
\begin{align}
({\cal C}^{-1})^{00}=&\frac{-1}{D}[\Gamma_N]^{0}\left\{ ([\Gamma_{N}]^{0})^{2} + 4 ( [{\rm H}\cdot {\rm H}] +[\widetilde{\rm H}\cdot \widetilde{\rm H}] )\right\}, \n
({\cal C}^{-1})^{i0}=&\frac{1}{D}{\big \{ }  ([\Gamma_{N}]^{0})^{2} [\Gamma_{N}]^{i} + 4 ([{\Gamma}_{N} \cdot {\rm H}] - [\widetilde{\Gamma}_{N} \cdot \widetilde{\rm H}] ) [{\rm H}]^{i} 
 -2[\Gamma_{N}]^{0} \epsilon^{ijk}[\Gamma_{N}]^{j}[{\rm H}]^{k} {\big \} }, \n
 (\widetilde{\cal C}^{-1})^{00}=&0, \n
(\widetilde{\cal C}^{-1})^{i0}=&\frac{1}{D}{\big \{ }  ([\Gamma_{N}]^{0})^{2} [\widetilde{\Gamma}_{N}]^{i} + 4 ([{\Gamma}_{N} \cdot {\rm H}] -[\widetilde{\Gamma}_{N} \cdot \widetilde{\rm H}] )  [\widetilde{\rm H}]^{i} \n
&\ \ \ \ \ \ -2[\Gamma_{N}]^{0} \epsilon^{ijk}[\Gamma_{N}]^{j}[\widetilde{\rm H}]^{k} -2[\Gamma_{N}]^{0} \epsilon^{ijk}[\widetilde{\Gamma}_{N}]^{j}[{\rm H}]^{k} {\big \} }, \label{calCinverse}
\end{align}
where $D$ is the determinant, 
\begin{align}
D\equiv & ([\Gamma_{N}]^{0})^{2} {\big \{ } ([ \Gamma_{N}]^{0})^{2}-[{ \Gamma}_{N} \cdot {\Gamma}_{N} ] +[\widetilde{ \Gamma}_{N} \cdot \widetilde{ \Gamma}_{N} ] + 4 ([{\rm H}\cdot \widetilde{\rm H}]+[{\rm  H}\cdot \widetilde{\rm  H}] ) { \big \} } \n
&\  - 4 {\big \{ }[{\Gamma}_{N} \cdot {\rm H}]+[\widetilde{\Gamma}_{N} \cdot \widetilde{\rm  H}]{\big \} }^2 \ .
\label{Determinant}
\end{align}
$[\ \cdot \ ]$ denotes a summation over $i=1,2,3$.
\subsection{CP-violation parameter $\varepsilon$}
In order to read the effective $CP$-violating parameter $\varepsilon$, we set $Y_L=0$ 
 and insert
(\ref{deltaYieldN}) into the kinetic equation of the lepton numbers (\ref{YieldL}),
\begin{align}
d_t Y_{L^{\alpha}}
= &{\rm Tr} {\Big [} i \Im (\Gamma_{\alpha}) \delta Y_{N}^{even} {\Big ]} 
+{\rm Tr}{\Big [} \Re (\widetilde{\Gamma}_{\alpha})\delta Y_{N}^{odd} {\Big ]} \n
=& 2 [\Gamma_{\alpha}]^{2} [\delta Y_{N}^{even}]^{2} + 2 \sum_{a=0,1,3} [\widetilde{\Gamma}_{\alpha}]^{a} [\delta Y_{N}^{odd}]^{a} \n
=& 2 \left\{ [ \Gamma_{\alpha}]^{2}({\cal C}^{-1})^{20} + [ \widetilde{\Gamma}_{\alpha}]^{1}(\widetilde{\cal C}^{-1})^{10}+[ \widetilde{\Gamma}_{\alpha} ]^{3}(\widetilde{\cal C}^{-1})^{30} \right\} \times  d_t Y_{N}^{eq}  \n
=& \frac{4[\Gamma_{N}]^0}{D}\epsilon^{ijk} \left\{ [ \Gamma_{\alpha}]^{i}[ \Gamma_{N}]^{j}[{\rm H}]^{k} + [ \widetilde{\Gamma}_{\alpha}]^{i}[ \Gamma_{N}]^{j}[\widetilde{\rm H}]^{k}+[ \widetilde{\Gamma}_{\alpha}]^{i}[ \widetilde{\Gamma}_{N}]^{j}[{\rm H}]^{k} \right\} \times (-  d_t Y_{N}^{eq} )
\label{YieldLadj}
\end{align}
The r.h.s. can be rewritten in terms of $2[\delta Y_N]^0 = {\rm Tr} ( \delta Y_N)$, 
which is the total RH neutrino number 
deviated from the local equilibrium.
Especially, neglecting the difference of helicity, we can write 
 the r.h.s. of (\ref{YieldLadj}) in terms of $[\delta Y_N^{even}]^0$  in (\ref{deltaYieldN})
as
\begin{align}
d_t Y_{L^{\alpha}}=& 2 \varepsilon^{\alpha}  [\Gamma_{N}]^0 [\delta Y_N^{even}]^0 \ . 
\label{defAveragedEpsilon} 
\end{align}
Here $[\Gamma_{N} ]^0$ is an averaged decay rate of RH neutrinos into charged lepton $\ell^{\alpha}$.
The CP-violating parameter $\varepsilon^\alpha$ defined by the coefficient is read as
\begin{align}
\varepsilon^{\alpha}&= \frac{2 \epsilon^{ijk} \left\{ [ \Gamma_{\alpha}]^{i}[ \Gamma_{N}]^{j}[{\rm H}]^{k} + [ \widetilde{\Gamma}_{\alpha}]^{i}[ \Gamma_{N}]^{j}[\widetilde{\rm H}]^{k}+[ \widetilde{\Gamma}_{\alpha}]^{i}[ \widetilde{\Gamma}_{N}]^{j}[{\rm H}]^{k} \right\} }{\left( ([\Gamma_{N}]^{0})^{2} + 4 ( [{\rm H}\cdot {\rm H}] +[\widetilde{\rm H}\cdot \widetilde{\rm H}] ) \right) [\Gamma_{N} ]^0} \n
&= - i
 \frac{{\rm tr} \left( \Gamma_\alpha \Gamma_N {\rm H} +\widetilde{\Gamma}_\alpha \Gamma_N
\widetilde{\rm H} + \widetilde{\Gamma}_\alpha \widetilde{\Gamma}_N {\rm H}\right)}
{ \left( ([\Gamma_{N}]^{0})^{2} + 4 ( [{\rm H}\cdot {\rm H}] +[\widetilde{\rm H}\cdot \widetilde{\rm H}] ) 
\right) [\Gamma_{N} ]^0}.
\label{epsilonformal} 
\end{align}
The result is valid when it is justified to replace $d_t Y_N$ by its equilibrium value $d_t Y_N^{eq}.$ 
Though our calculation fixes the flavour basis in which the Majorana masses are diagonal,
the final form is written in a flavour covariant way.
The above definition of $\varepsilon$ is appropriate since the numerator of the ordinary definition 
\be
\varepsilon
&\equiv& \frac{\Gamma_{N \to \ell \phi}-\Gamma_{N \to \overline{\ell} \overline{\phi}}}{\Gamma_{N \to \ell \phi}+\Gamma_{N \to \overline{\ell} \overline{\phi}}}
\ee
is replaced by $d_t Y_L/2[\delta Y_N]^0$ while the denominator is approximated by $ \Gamma_N.$
\subsection{Explicit forms of $\delta Y_N$}
In this section, we use explicit forms of various quantities to rewrite the formal
expression (\ref{epsilonformal}) in a more familiar form.

${\rm H}$  ($\widetilde{\rm H}$) is the helicity even (odd)
 part of the mass (with thermal corrections included)
and given in (\ref{H}). $\widetilde{\rm H}$ has an $a=2$ component only.
For ${\rm H}$, $a=0$ component is the total mass and decouples from the 
equation. $a=3$ component of H gives the mass difference
\begin{align}
2 [{\rm H}]^{3}&= \frac{\xi_{0}}{sY_{N}^{eq}} (M_{1} -M_{2}) +\cdots
\label{H3}
\end{align}
where
\begin{align}
\xi_{0} \equiv 2 M\int \frac{dq^3}{(2\pi)^3}\frac{1}{\omega_{q}} f^{eq}_{N q} .
\end{align}
The $\cdots$ in $[{\rm H}]^{3}$ represents finite temperature (and density) corrections to the RH neutrino potential.
Off-diagonal components $[{\rm H}]^{1}$ and $[\widetilde{\rm H}]^{2}$ represent kinetic mixing
induced  by the thermal effects, and can be removed  by flavour rotation
at each time. Unitary matrix diagonalizing the mass matrix is time dependent, but in the following
analysis, we neglect time-dependence of the thermal mass and mixing. 
If we neglect the statistical effects,
the coefficient in $[H]^3$ is given by $(\xi_{0}/sY_{N}^{eq})=K_1(M/T)/K_2(M/T)$.
At low temperature $T\ll M$ it approaches $(\xi_{0}/sY_{N}^{eq}) \to 1$
while at high temperature $T\gg M$, it behaves as $(\xi_{0}/sY_{N}^{eq}) \sim M/(2T)$.

$\Gamma_N$ comes from the self-energy diagrams of RH neutrinos, and
contains information of (inverse) decay or scattering of RH neutrinos. 
We  decompose $\Gamma_N$ into $\Gamma_\alpha$ by fixing the  flavour $\alpha$ 
of lepton $\ell^\alpha$ in  the final state. Only the real part appears in the KB equation. 
From (\ref{GammaN}), we can decompose $\Gamma_N$ in the model (\ref{Lint}) as
\begin{align}
\Gamma_N &= \frac{\xi}{sY_{N}^{eq}} \frac{\Re(h^{\dag}h) M}{8\pi} + 
\Gamma_{N}^{\rm scatt} +\Gamma_{N}^{\rm vertex} \ , \label{D+S+V}
\end{align}
where
\begin{align}
\xi  \equiv & 32\pi \left( M -\frac{m_{\phi}^{2}-m_{\ell}^{2}}{M} \right)\int d\Pi_{N \ell^{\alpha} \phi}f^{eq}_{N q} (1-f^{eq}_{\ell p}+f^{eq}_{\phi k}) \ . 
\end{align}
$\Gamma_\alpha$ is  a partial decay width that RH neutrino decays into $\ell^\alpha$.
At the leading order, it is given by replacing  $(h^\dagger h)_{ij}$ in (\ref{D+S+V})
by $(h^\dagger_{i\alpha}h_{\alpha j})$ (no summation over $\alpha$).

The first term of $\Gamma_N$ is the decay amplitude at the tree level
and if we neglect the statistical effects and the thermal mass of the Higgs and lepton,
$\xi$ coincides with $\xi_0$, and approaches
 \be
 (\xi/sY_{N}^{eq})=(\xi_{0}/sY_{N}^{eq}) \rightarrow M/(2T)
 \ee
at high temperature.
$\Gamma_{N}^{\rm scatt}$ are corrections to the 
decay rate from scattering with the top quarks or gauge particles in the thermal media. 
$\Gamma_{N}^{\rm vertex}$ are  corrections to the vertex diagram.
It is negligible compared to the first term. 
In the  resonant leptogenesis, 
the direct CP violating parameter associated with an interference between 
the tree and the  vertex correction  can be neglected 
compared to the indirect CP violation   through the flavour oscillation.
Then the relations 
 $[\Gamma_{N}]^2=[\widetilde{\Gamma}_{N}]^{0,1,3}=0$ hold. (See footnote \ref{footnoteRI}.)

In order to simplify the notation, we write
\begin{align}
\left( \Gamma_N \right)_{ij}
&= \frac{\xi_0}{sY_{N}^{eq}}  \Gamma_{ij}^{\rm eff} \   , \ \   
 ( \widetilde{\Gamma}_N )_{ij}
= \frac{\xi_0}{sY_{N}^{eq}} \widetilde{\Gamma}_{ij}^{\rm eff}  \label{GammaEff}
\end{align}
where $ \Gamma^{\rm eff}_{ij} $ and $ \widetilde{\Gamma}^{\rm eff}_{ij} $ are  effective decay rates including not only thermal effects but also scattering contributions. 
If interactions do not change the flavour structure, the effective decay matrix
is written as
\begin{align}
\Gamma_{ij}^{\rm eff}  &= (1+\alpha) M\frac{\Re (h^{\dag}h)_{ij}}{8\pi}  \ , \ \ 
\widetilde{\Gamma}_{ij}^{\rm eff}  = \tilde{\alpha} M\frac{i \Im (h^{\dag}h)_{ij}}{8\pi} \ . \label{atilde}
\end{align}
for $a=1,2,3$ component.
  Furthermore, if we consider 
 flavour independent
interactions such as  $B-L$ gauge interaction of RH neutrinos,
an additional contribution is added to  $a=0$ component $[\Gamma_N]^0$.
In the following, we neglect this contribution for simplicity.
When we neglect thermal effects and scattering contributions, $\alpha$ and $\widetilde{\alpha}$ vanish and diagonal components of $\Gamma_{ii}^{\rm eff}$  are reduced to the tree-level vacuum decay rate 
$\Gamma_{i}^{\rm vac}\equiv (h^{\dag}h)_{ii}M/(8\pi)$.
In the following we write $\Gamma_i = \Gamma^{\rm eff}_{ii}$ as a decay rate including the above corrections.

Using these quantities of ${\rm H}$ and $\Gamma_N$, we can 
express each component of the inverse matrix ${\cal C}^{-1}$ in terms of
masses $M_i$ and  decay rates $\Gamma_i$.
The explicit forms are written in Appendix B.

By using the explicit forms of ${\cal C}^{-1}$ in Appendix B, 
we can write down each component of $\delta Y$ as follows.
First, the diagonal components of 
$\delta Y_{N}^{even}$ ($a=0,3$) are given by
\begin{align}
[\delta Y_N^{even}]^0=&-\frac{d_t Y_N^{eq}}{\xi_{0} /(sY_N^{eq})} \frac{\Gamma_{1}+
\Gamma_{2}}{2\Gamma_{1}\Gamma_{2}}  U , 
\end{align}
\begin{align}
[\delta Y_N^{even}]^3=&-\frac{d_t Y_N^{eq}}{\xi_{0} /(sY_N^{eq})} \frac{-\Gamma_{1}+\Gamma_{2}}{2\Gamma_{1}\Gamma_{2}}  U , 
\end{align}
where 
\begin{align}
U\equiv \frac{(M_{1}^2-M_{2}^{2})^2 +M^2(\Gamma_{1}+\Gamma_{2})^2 }{(M_{1}^2-M_{2}^{2})^2 + M^2 (\Gamma_1+\Gamma_2)^2 X
} \ .
\end{align}   
and 
\be
X= 
\frac{{\rm det}[\Re (h^{\dag}h)](1+\alpha)^2 -(\widetilde{\alpha}\Im (h^{\dag}h) )^2}{(h^{\dag}h)_{11}(h^{\dag}h)_{22} (1+\alpha)^2} \ .
\ee
$[\delta Y_{N}^{even} ]^0$ gives an  averaged number of the RH neutrinos deviated from the local equilibrium.
Equivalently, $ii$-component of the matrix $\delta Y_N^{even}$ is given by 
\begin{align}
\left( \delta Y_{N}^{even}\right)_{ii}
=&[\delta Y_N^{even}]^0 \pm [\delta Y_N^{even}]^3=-\frac{d_t Y_N^{eq}}{\xi_{0} /(sY_N^{eq})} \frac{U}{\Gamma_{i}}  
\end{align}
where $\pm$ represents $i=1,2$ respectively.

Off-diagonal components can be similarly obtained.
The real part $a=1$ and the imaginary part $a=2$  of $\delta Y_N^{even}$  are given by
\begin{align}
[\delta Y_N^{even}]^1&=\Re \delta Y_{N 12}^{even} =
-2(1+\alpha) \Re [h^{\dag} h]_{12}(\Gamma_{1}+\Gamma_{2})M V
[\delta Y_N^{even}]^0 , \\
[\delta Y_N^{even}]^2&=
-\Im \delta Y_{N 12}^{even} = 
-2(1+\alpha)\Re [h^{\dag} h]_{12}(M_1 ^2-M_2^2) V [\delta Y_N^{even}]^0  .
\label{dYoffdiag}
\end{align}
For $\delta Y_N^{odd}$, we have
\begin{align}
[\delta Y_N^{odd}]^1&=\Re \delta Y_{N 12}^{even} =
2\widetilde{\alpha} \Im [h^{\dag} h]_{12}(\Gamma_{1}+\Gamma_{2})M V
[\delta Y_N^{even}]^0 , \\
[\delta Y_N^{odd}]^2&=-\Im \delta Y_{N 12}^{even} =
-2\widetilde{\alpha}\Im [h^{\dag} h]_{12}(M_1 ^2-M_2^2) V [\delta Y_N^{even}]^0  .
\label{dYoffdiag}
\end{align}
Here we defined
\be
V \equiv \frac{M^2/(8\pi)}{(M_{1}^2-M_{2}^{2})^2 +M^2(\Gamma_{1}+\Gamma_{2})^2 } .
\ee
$[\delta Y_N^{even}]^2$ and $[\delta Y_N^{odd}]^1$ give the CP violating parameter $\varepsilon$.
It is given in a simplified case in the next section.

We comment on a situation 
when ${\rm det}[\Re (h^{\dag}h)]$ becomes small. (For simplicity we set $\tilde{\alpha}$=0.)
Then $X$ and accordingly $[\delta Y_N^{even}]^0$ is largely enhanced.
The situation corresponds to a case that an effective decay rate (cf.(\ref{Yieldeven})) is small.
Especially when the mass difference vanishes $M_1=M_2$, it diverges at ${\rm det}[\Re (h^{\dag}h)]=0$, 
namely when ${\rm det}C=0$.  
In such a situation, the deviation of RH neutrino number density becomes large and the 
assumption of our investigation, smallness of the deviation from local equilibrium, becomes invalid.
\subsection{CP violating parameter $\epsilon$ \label{SecCPsimple} when $\widetilde{\cal C}=0$}
Finally we write the formal expression of (\ref{epsilonformal}) in a more familiar
form by introducing further simplifications. 
We neglect the thermal mass of leptons and drop the Pauli blocking terms.
Then the helicity odd part of $\gamma_h^{\ell \phi}$ disappears as explained in 
(\ref{hoddintegral}) and  the off-diagonal components 
$\widetilde{\cal C} $ connecting the CP even and odd parts in $\delta Y$ vanish.
Furthermore we use the vacuum value of $\Gamma_N \ (\alpha=\widetilde{\alpha}=0). $
Then, by using explicit forms of ${\rm H}$ in (\ref{H3}) and $\Gamma_N$ in (\ref{GammaEff})
with $\Gamma_{ij}^{\rm eff}=\Gamma_{ij}^{\rm vac}$,
the CP-violating parameter $\varepsilon^\alpha$  is given by
\begin{align}
\varepsilon^{\alpha}&= 
\frac{2 \epsilon^{ijk} [\Gamma_{\alpha}]^{i}[ \Gamma_{N}]^{j}[{\rm H}]^{k}  }{([\Gamma_{N}]^{0})^{2} + 4 [{\rm H}\cdot {\rm H}] } \n
&=\frac{2\Re (h^{\dag} h)_{12} \Im (h_{1\alpha}^{\dag} h_{\alpha 2})}{((h^{\dag} h)_{11}+(h^{\dag} h)_{22})^2/4}\frac{ (M_1 ^2-M_2^2)M (\Gamma_{1}+\Gamma_{2})/2}{(M_{1}^2-M_{2}^{2})^2 +
M^2 (\Gamma_{1}+\Gamma_{2})^2} \ .
\label{epsilonfam}
\end{align}
This CP violating parameter has the regulator $M^2 (\Gamma_{1}+\Gamma_{2})^2$
which is consistent with our previous result \cite{ISY}.
In the previous analysis we obtained the same result under an assumption that the off-diagonal 
Yukawa couplings are smaller than the diagonal ones. In the present analysis, we 
do not use such a condition, and  take effects of  coherent flavour oscillation fully into account.
The decay widths $\Gamma_i^{\rm eff}$ are determined by the effective decay width (\ref{GammaEff}),
which are obtained from the 1PI self-energy diagrams $\Pi$ by cutting the diagrams and putting 
external lines on mass-shell.

Finally we note that
we can decompose the r.h.s. of (\ref{defAveragedEpsilon}) 
into $N_i$ ($i=1,2$) as
\begin{align}
d_t Y_{L^{\alpha}}
=&\sum_{i=1,2} \varepsilon_{i}^{\alpha}   \left( \Gamma_{N}\right)_{ii}  \left( \delta Y_{N}^{even} \right)_{ii} 
\label{defAveragedEpsiloni}
\end{align}
where we define the CP violating parameter of each $N_i$ as
\begin{align}
\varepsilon_{i}^{\alpha}&=\frac{2\Re (h^{\dag} h)_{12} \Im (h_{1\alpha}^{\dag} h_{\alpha 2})}{(h^{\dag} h)_{11}(h^{\dag} h)_{22}}\frac{ (M_1 ^2-M_2^2)M \Gamma_{j(\ne i)}}{(M_{1}^2-M_{2}^{2})^2 
+M^2 (\Gamma_{1}+\Gamma_{2})^2}  \ .
\label{epsiloni}
\end{align}
When $i=1$, $j$ takes $2,$ and vice-versa. 
Such a separation into a different flavour of RH neutrinos
is, of course,  valid only when the off-diagonal component $(h^{\dag}h)$ is smaller than 
the diagonal one.
The numerator of the first factor can be rewritten as 
\begin{align}
2\Re (h^{\dag} h)_{12} \Im (h_{1\alpha}^{\dag} h_{\alpha 2})
= \Im [ (h^{\dag} h)_{12} (h_{1 \alpha}^{\dag} h_{\alpha 2})]+\Im [ (h^{\dag} h)_{21} (h_{1 \alpha}^{\dag} h_{\alpha 2})]
\end{align}
which gives a consistent result with \cite{Covi:1996wh}.

\section{Summary  \label{sec-Summary}}
In the paper, we solved the KB equation without assuming that  the off-diagonal component of 
the Yukawa couplings are small compared to the diagonal ones.
In order to solve it, we first derive the kinetic equation for the density matrix. 
The differential equation can be reduced to a linear equation if the background 
is slowly changing and the deviation of the distribution function from local equilibrium
is small. Then the density matrix of RH neutrino
 can be solved in terms of the time variation of the equilibrium
distribution function and the generated lepton asymmetry.
Its off-diagonal component 
determines the CP violating parameter $\varepsilon.$
It is resonantly enhanced due to the almost degenerate Majorana masses
and the regulator of $\varepsilon$ is given by $R_{ij}=M_i \Gamma_i+M_j \Gamma_j$.
In the 2PI formalism, the decay width $\Gamma_i$ is given by the imaginary part of 
the self-energy function of the RH neutrinos. 
In addition to the loop corrections of the vertex functions, scattering effects with 
particles in medium are contained. 
The effect of coherent oscillation is fully taken into account by considering the 
density matrix formalism.

The derivation of the kinetic equation of the density matrix from the KB equation
is based on an assumption that the distribution function is not far from 
the local equilibrium. It will be interesting to obtain the kinetic equation when the system
is far from equilibrium. We want to come back to this problem in near future.

\section*{Note added}
During the final stage of writing the manuscript, an interesting  paper \cite{Pila2014} appeared.
In the paper, the authors derived the kinetic equation of density matrix based on the Hamiltonian
approach, and solve the equation to obtain $\delta Y_N^{even}$ in the flavour covariant way. 
The result is consistent with ours but the interpretation of the CP violating parameter seems to be different.
Also, in  \cite{Pila2014},
the one-loop resummed effective Yukawa coupling is used to define decay and inverse-decay
amplitudes ($\Gamma_N$ in our notation), in which the effect of coherent oscillation is included
in their analysis. 
In our approach based on the 2PI formalism, $\Gamma_N$ comes from 1PI self-energies
and the effect of coherent oscillation is not contained.
The indirect CP violating parameter $\epsilon$ generated by
resummation of RH neutrino propagators 
is taken into account by considering  the multi-flavour formulation of density matrix.

\section*{Acknowledgements}
We would like to thank Masato Yamanaka for discussions.
This work is supported
in part by Grant-in-Aid for Scientific Research
(No. 23244057, 23540329) 
from the Japan Society for the Promotion of Science, and
``The Center for the Promotion of Integrated Sciences (CPIS)'' of Sokendai.
\appendix
\section{ Derivation of the kinetic term $d_t f_N$ \label{AppA} }
\setcounter{equation}{0}
In this appendix, we show how the the kinetic term in 
(\ref{DMevoeq}) 
$-i d_t f_{N, h,q}$
 is derived from the l.h.s. in (\ref{frequencyrepr}):
\be
& -i{\rm tr}{\Big [} P_h 
{\Big (}\diamondsuit \left\{ \gamma^{0}q_0-\frac{{\bold q}\cdot {\boldsymbol \gamma}}{a}-\hat{M}-\Pi^{eq}_R \right\} \left\{ if  \right\} G^{eq}_{\rho} 
- \Pi^{eq}_{\rho}
 \diamondsuit \left\{ if\right\} \left\{G_{A}^{eq}\right\} 
\n
&  -
G^{eq}_{\rho} \diamondsuit \left\{ if  \right\} \left\{ \gamma^{0}q_0-\frac{{\bold q}\cdot {\boldsymbol \gamma}}{a}-\hat{M}-\Pi^{eq}_A \right\}  
+ \diamondsuit \left\{G_{R}^{eq} \right\} \left\{ if\right\} \Pi^{eq}_{\rho} {\Big )} {\Big ]}  \ .
\label{kinetictermKB}
\ee

First we look at the leading term. 
For simplicity,  we drop the self-energy correction $\Pi_R^{eq}$. Then we have
\begin{align}
&i{\rm tr}\left[ P_{ h} \sum_{ h'} \diamondsuit \left\{ \gamma^{0}q_0-\frac{{\bold q}\cdot {\boldsymbol \gamma}}{a}-\hat{M} \right\} \left\{ if^{eq}_{ h'}  \right\} (\Slash{q}+M)P_{ h'} \right] 
\frac{\Gamma_{a}}{((q_0 -\omega_{q})^2 +\Gamma_{q}^2 /4)}
\n
&= q_0 \left(\partial_{X} f^{eq}_{ h}(q_0,X) -\frac{H|{\bold q}|^2}{q_0 a^2} \partial_{q_0} f^{eq}_{ h}(q_0,X) \right) \frac{\Gamma_{a}}{((q_0 -\omega_{q})^2 +\Gamma_{q}^2 /4)}
\label{totalderivative} 
\end{align}
If we set  $q_0 = \omega_q$,   two terms in the bracket give a total derivative 
\be
d_t = (\partial_{t} T)\partial_{T} + (\partial_{t} \omega_{q})\partial_{\omega_q}
\ee
 of the on-shell Fermi distribution function $f^{eq}_{{ h}q}\equiv f^{eq}(t,\omega_{q}(t))$ in equilibrium. 
But the propagator has a Lorentz type structure and $q_0$ is extended
around the position of the pole $q_0=\omega_q$.

We then take an effect of the remaining terms in (\ref{kinetictermKB}).
These terms can be rewritten as
\begin{align}
&i{\rm tr} \left\{ P_{ h} \Pi_{\rho}\diamondsuit \left\{ if^{eq}\right\} \left\{G_{A}\right\} - P_{ h} \diamondsuit  \left\{G_{R}\right\} \left\{ if^{eq} \right\} \Pi_{\rho} \right\}  \n
&=i{\rm tr} \left\{ P_{ h} \Pi_{\rho} G_{A} \diamondsuit \left\{ if^{eq} \right\} \left\{G_{A}^{-1}\right\}G_{A} - P_{\sf h} G_{R} \diamondsuit  \left\{G_{R}^{-1}\right\} \left\{ if^{eq} \right\} G_{R} \Pi_{\rho} \right\}  \n
&\simeq -{\rm tr} \left\{ P_{h}  \diamondsuit \left\{ \gamma^{0}q_0-\frac{{\bold q}\cdot {\boldsymbol \gamma}}{a}-\hat{M} \right\} \left\{ f^{eq} \right\} \left(G_{R} \Pi_{\rho} G_{R}+ G_{A} \Pi_{\rho} G_{A} \right) \right\} \ .
\label{broaden}
\end{align}
In the first equality, we have used the relation $\diamondsuit \{f\} \{A\} = -\diamondsuit \{A\} \{f\} $ and $\diamondsuit \{f\} \{A\}=A \diamondsuit \{f\} \{A^{-1}\} A$ for a given matrix $A$.
In the second line, we have used $G_{R/A}^{-1}=-(\Slash{q}-\hat{M}-\Pi_{R/A})$ and droped next-to-leading order contributions $\Pi_{R,A}$.

Using (\ref{broaden}),  four terms in (\ref{kinetictermKB}) are combined to become
\begin{align}
&2{\rm tr} \left\{ P_{ h}  \diamondsuit \left\{ \gamma^{0}q_0-\frac{{\bold q}\cdot {\boldsymbol \gamma}}{a}-\hat{M} \right\} \left\{ f \right\} G_{\rho} \right\} \n 
&-{\rm tr} \left\{ P_{ h}  \diamondsuit \left\{ \gamma^{0}q_0-\frac{{\bold q}\cdot {\boldsymbol \gamma}}{a}-\hat{M} \right\} \left\{ f^{eq} \right\} \left(G_{R} \Pi_{\rho} G_{R} + G_{A} \Pi_{\rho} G_{A} \right) \right\} \n 
&\simeq \left(\partial_{X} f_{h}(q_0,X) -\frac{H|{\bold q}|^2}{q_0 a^2} \partial_{q_0} f_{ h}(q_0,X) \right) \times (-i) \n
&\ \ \ \ \times \left( \frac{\Gamma_q}{(q_0 -\omega_{q})^2 +\Gamma_{q}^2 /4} - \frac{\Gamma_q  \left( q_{0} -\omega_{q}-i\Gamma_{q}/2 \right)^2}{2((q_0 -\omega_{q})^2 +\Gamma_{q}^2 /4 )^2} - \frac{\Gamma_q  \left( q_{0} -\omega_{q}+i\Gamma_{q}/2 \right)^2}{2((q_0 -\omega_{q})^2 +\Gamma_{q}^2 /4 )^2} \right) \n
&= -i \left(\partial_{X} f_{ h}(q_0,X) -\frac{H|{\bold q}|^2}{q_0 a^2} \partial_{q_0} f_{ h}(q_0,X) \right) \times \frac{\Gamma_{q}^3/2}{((q_0 -\omega_{q})^2 +\Gamma_{q}^2 /4)^2} 
\label{sharpspectrum}
\end{align}
around the position of the pole $q_0=\omega_q$.
Here, we used the approximate form $\Pi_{\rho} \sim \Slash{q} \times (-i\omega_{q_0} \Gamma_{q}/M^2)$ and dropped higher order  terms with respect to  $(q_0 - \omega_{q})$. 
Hence, the original Lorentz type distribution becomes to have a sharper spectrum 
after adding the higher order terms in the KB equation.
Namely, the term $\Gamma_{q}^3/2/((q_0 -\omega_{q})^2 +\Gamma_{q}^2 /4)^2$ 
approaches Dirac delta function 
 $2\pi \delta(q_0 -\omega_{q}) $ much faster than 
 the usual Lorentz type form $\Gamma_{a}/((q_0 -\omega_{q})^2 +\Gamma_{q}^2 /4)$
  in the limit $\Gamma_{q} \to 0$ \cite{MorozovRopke}.
  
In this appendix, we considered a single flavour case in order to see that the distribution function is sharpened as above. The effect of flavour mixing due to the second term in (\ref{sharpspectrum}) may change the flavour structure in the l.h.s. of the kinetic equation.
We want to come back to this interesting issue in future.
\section{Appendix B: Explicit forms of ${\cal C}^{-1}$ and $\tilde{\cal C}^{-1}$  \label{AppB}} 
\setcounter{equation}{0}
\begin{align}
({\cal C}^{-1})^{a0}&=\frac{-1}{D}\left( \begin{matrix} [\Gamma_N]^{0}\left\{ ([\Gamma_{N}]^{0})^{2} + (2[{\rm H}]^3)^2\right\} \\ -([\Gamma_N]^{0})^2 [\Gamma_N]^{1} \\ -2 [{\rm H}]^3 [\Gamma_N]^{0} [\Gamma_N]^{1} \\ -[\Gamma_N]^{3}\left\{ ([\Gamma_{N}]^{0})^{2} +  (2[{\rm H}]^3)^2\right\} \end{matrix} \right)^{a} \n
&= \frac{-\xi_{0}^{3}}{D (sY_{N}^{eq})^{3}}\left( \begin{matrix} [\Gamma]^{0}\left\{ ( [\Gamma]^{0})^{2} + (M_1 -M_2)^2\right\} \\ -([\Gamma]^{0})^2 [\Gamma]^{1} \\ - (M_1-M_2) [\Gamma]^{0} [ \Gamma]^{1} \\ -[\Gamma]^{3}\left\{ ( [\Gamma]^{0})^{2} + (M_1 -M_2)^2 \right\} \end{matrix} \right)^{a}, \  \n
(\widetilde{\cal{C}}^{-1})^{a0}&=\frac{-1}{D}\left( \begin{matrix} 0 \\  +2 [{\rm H}]^3 [\Gamma_N]^{0} [\widetilde{\Gamma}_N]^{2} \\ -([\Gamma_N]^{0})^2 [\widetilde{\Gamma}_N]^{2} \\ 0 
\end{matrix} \right)^{a}
= \frac{-\xi_{0}^3 }{D (sY_{N}^{eq})^{3}}\left( \begin{matrix} 0 \\  +(M_1 -M_2) [ \Gamma]^{0} [\widetilde{\Gamma}]^{2} \\ - ([\Gamma]^{0})^2 [\widetilde{\Gamma}]^{2} \\ 0 
\end{matrix} \right)^{a} \  ,  
\end{align}
where determinant  $D$ is given by
\begin{align}
D&=\left\{ ([\Gamma_{N}]^{0})^{2} - ([\Gamma_{N}]^{3})^{2} \right\} \left[ (2[{\rm H}]^{3})^2 +  ([\Gamma_{N}]^{0})^{2} \frac{([\Gamma_{N}]^{0})^{2}-[\Gamma_{N}\cdot \Gamma_{N}]-[\widetilde{\Gamma}_{N}\cdot \widetilde{\Gamma}_{N}]}{([\Gamma_{N}]^{0})^{2}-([\Gamma_{N}]^{3})^{2}} \right] \n
& = \frac{\xi_{0}^4 }{(sY_{N}^{eq})^4} \Gamma_{1} \Gamma_{2}  \left[ (M_1 -M_2)^2 +  ([\Gamma]^{0})^{2} \frac{{\rm det}\{ \Gamma  \}- ([\widetilde{\Gamma}]^2)^2}{\Gamma_{1} \Gamma_{2}} \right] .
\end{align}

\end{document}